\documentclass[aps,prl,preprint
,
superscriptaddress,floatfix, 
longbibliography]{revtex4-2}
\usepackage[version=3]{mhchem}

\usepackage{amsmath}
\usepackage{amsfonts} 
\usepackage{graphicx}
\usepackage{verbatim}
\usepackage{natbib}

\usepackage{booktabs}

\usepackage{multirow}

\begin{document}

\title{Light Control over Chirality Selective Functionalization of Substrate Supported Carbon Nanotubes}

\author{Georgy Gordeev}
\email{gordeev@zedat.fu-berlin.de}
\affiliation{Department of Physics, Freie Universität Berlin, 14195 Berlin, Germany}
\affiliation{Department of Physics and Materials Science, University of Luxembourg, L-4422 Belvaux, Luxembourg}


\author{Thomas Rosenkranz}
\affiliation{Institute of Nanotechnology, Karlsruhe Institute of Technology, 76021 Karlsruhe, Germany}
\affiliation{Institute of Materials Science, Technische Universität Darmstadt, 64287 Darmstadt, Germany}

\author{Frank Hennrich}
\affiliation{Institute of Nanotechnology, Karlsruhe Institute of Technology, 76021 Karlsruhe, Germany}
\affiliation{Institute of Quantum Materials and Technologies, Karlsruhe Institute of Technology, 76021 Karlsruhe, Germany}

\author{Stephanie Reich}

\affiliation{Department of Physics, Freie Universität Berlin, 14195 Berlin, Germany}

\author{Ralph Krupke}
\email{ralph.krupke@kit.edu}
\affiliation{Institute of Nanotechnology, Karlsruhe Institute of Technology, 76021 Karlsruhe, Germany}

\affiliation{Institute of Quantum Materials and Technologies, Karlsruhe Institute of Technology, 76021 Karlsruhe, Germany}

\affiliation{Institute of Materials Science, Technische Universität Darmstadt, 64287 Darmstadt, Germany}
\date{\today}

\begin{abstract}
  Diazonium reactions with carbon nanotubes form optical $sp^3$ defects that can be used in optical and electrical circuits. We investigate a direct on-device reaction supported by confined laser irradiation and present a technique where an arbitrary carbon nanotube can be preferentially functionalized within a device by matching the light frequency with its transition energy. An exemplary reaction was carried out between (9,7) nanotube and 4-bromobenzenediazonium tetrafluoroborate. The substrate supported nanotubes of multiple semiconducting chiralities were locally exposed to laser light while monitoring the reaction kinetics \textit{in-situ} via Raman spectroscopy. The chiral selectivity of the reaction was confirmed by resonant Raman spectroscopy, reporting a 10 meV $E_{22}$ transition energy red-shift only of the targeted species. We further demonstrated this method on a single tube (9,7) electroluminescent device and show a 25 meV red-shifted emission of the ground state $E_{11}$ compared to the emission from the pristine tubes.
  \includegraphics[width=11cm]{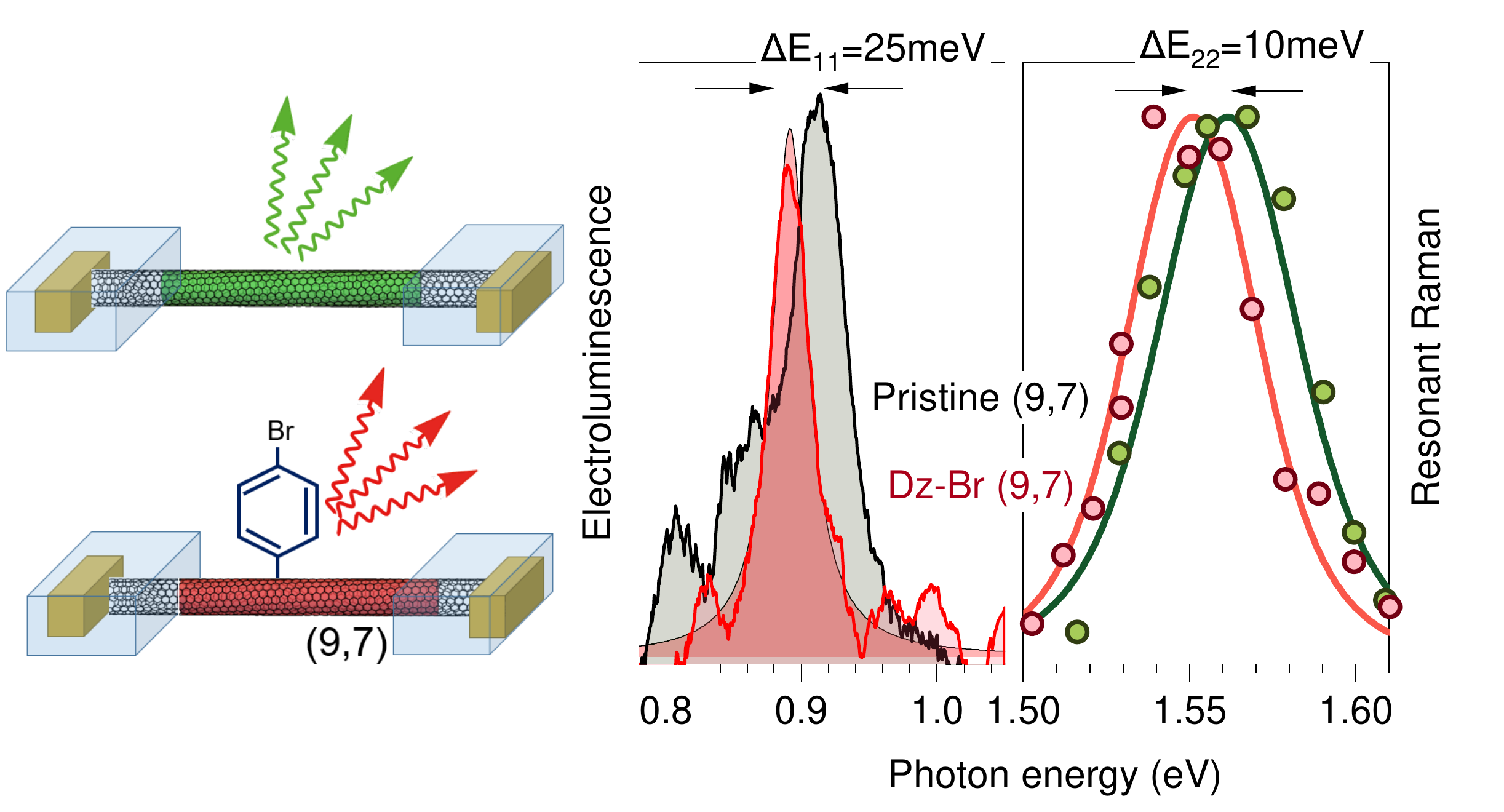}
  
  \textbf{Key words: carbon nanotube, functionalization, in-situ reaction, defects engineering, Raman spectroscopy, photoinduced chemistry}
\end{abstract}

\maketitle
\newpage

\section{Introduction}
Single-walled carbon nanotubes (SWCNTs) are unique one-dimensional crystals with outstanding mechanical, optical, and electrical properties.\cite{Thomsen2007, He2018} The properties of the CNTs could be improved by introducing functional groups onto the CNTs surface. Such functionalization can improve intrinsic CNTs properties\cite{Sun2002}, merge CNTs properties  with those of functional moiety \cite{Ernst2012} or even create new optical states via defect engineering \cite{Piao2013,He2019}. The most robust approach is covalent functionalization where a chemical bond is created between functional moieties and the CNT. Recently tremendous attention has been paid to covalent functionalization techniques with aryl-\cite{Piao2013} alkane-\cite{Kwon2016} and epoxide\cite{Ma2014} groups, because they enable precise control over the defect type and yield novel properties. For fully harvesting such properties in devices, the functionalized CNTs require a reliable integration into electrical and photonic circuits\cite{Luo2017,Mueller2010,Khasminskaya2016}. There are two possible pathways, either the CNT is functionalized before or after integration. While the first method is applicable for a standard solution processed material, the lateral position of the functional group is not well controlled and might occur not at the center of a cavity or between electrods but somewhere close to a contact, where optical emission can be suppressed. An alternative approach is to functionalize CNTs directly on a device. For this novel approach a functionalization reaction must be developed for substrate supported configurations.

The reaction between CNTs and diazonium salts (Dz)\cite{Strano2003,Usrey2005,Mohamed2015} is sensitive to visible light.\cite{Saha2018,Powell2016,Powell2017,Kwon2016}  Aryl ring functionalization first occurs in metallic CNTs, which has been used to enrich CNTs of electronic type\cite{Strano2003,Kim2007}. In semiconducting CNTs the reaction is inherently slow and can take weeks to succeed, but when light drives the reaction it accelerates to a just few hours \cite{Powell2016}. This activation process is more effective with photons whose energy matches absorption\cite{Powell2016,Powell2017} occurring at optical transitions $E_{11}$, $E_{22}$, etc.\cite{Thomsen2007} The transition energies of the CNT are determined by the (n,m) chirality, therefore offering a pathway for chiral selectivity. Light driven reactions were so far only performed in bulk aqueous,\cite{Piao2013,Kwon2016,Powell2016} organic solutions,\cite{Berger2019} and dried films.\cite{Huang2020} The questions remains whether an on-device reaction occurs at a liquid-solid phase boundary, given the limited amount of access points, and whether the reaction at the surface is still light sensitive and chirality selective. Another challenge is that reaction monitoring by the optical signatures of defect states may be inaccessible in devices. We  therefore need to find visible alternatives, such as electrical performance, vibrational spectroscopy, or changes in the energies of the delocalized excitons.

The monitoring of the reaction \textit{in-situ} is crucial for understanding the reaction kinetics and the physical effects of functionalization such as strain or doping. The evolution of Raman spectra can be used to monitor the reaction \textit{in-situ}. The covalent bond formation is accompanied by the $sp^3$ distortion in the $sp^2$ hybridized CNT lattice. This distortion breaks the translational symmetry and introduces a defect state \cite{Maultzsch2001}. Thus the selection rules for momentum forbidden Raman scattering processes change. The defect induced modes occur $\sim$ 1300-1350 $cm^{-1}$ in most $sp^2$ carbons, including graphite\cite{Thomsen2000}, graphene\cite{Cancado2011,Cancado2017}, and CNTs\cite{Maultzsch2001}. The D mode has been widely used in functionalized materials,\cite{Mohamed2015,Gordeev2016a} including aryl ring functionalization\cite{Kim2007}. The increase of the D mode can therefore serve as a scale of the reaction product, since it is proportional to defects concentration. An additional doping effect, imposed by a functional group, can also be detected via Raman\cite{Grimm2017}. The expected changes in the G mode positions of semiconducting CNTs are on the order of 0.5 cm$^{-1}$,\cite{Das2009} and are difficult to analyze \textit{ex-situ}.

Our manuscript consists of three sections; in the first we study the reaction kinetics and light sensitivity on a mixed chirality sample. From all chiralities we target (9,7) CNT with a 1.58 eV laser and monitor \textit{in-situ} evolution of the Raman modes. We determine the influence of the photon flux on the reaction rates and products. Further, we investigate the functionalization spot and find the defects exclusively in the areas irradiated by light. In the second section we investigate chiral selectivity. The transition energy shifts of three CNT species are investigated by resonance Raman spectroscopy. The red-shift is only found in the targeted (9,7) CNT. With this we establish an additional transition energy-based metric of functionalization. In section three, we apply this metric for a single-tube electroluminescence device\cite{Gaulke2020}. The functionalization occurs after CNT integration onto a device and its spectral output indicates a successful reaction together with a change of electrical transport characteristics.

\section{Experimental Methods}
The very first step is the preparation of the \textit{semiconducting CNTs suspension}. The nanotubes need to be purified first, in order to remove all metallic species. The HiPco-CNT-suspensions in toluene were prepared using the polymer wrapping technique \cite{Sturzl2009}. The Poly(9,9-di-n-octylfluorenyl-2,7-diyl) (PFO) provided from Sigma Aldrich was used as a polymer surfactant. The excess polymer was removed with the help of size-exclusion chromatography (SEC).\cite{Hennrich2016} The suspensions contain $\mu$-long s-CNTs with almost no m-CNTs and low polymer concentration.

We found that With \textit{laser-assisted functionalization} (LAF) it is possible to select the desired chirality and use scattered light for monitoring reaction \textit{in-situ}.  The DzBr salt was purchased from Sigma Aldrich. With a molecular weight of 270.82 g/mol and 96 \% purity yielding a concentration of 3.5 $\mu$mol/g when 1 mg is dissolved in 1 g of water. The freshly prepared solution has a pH-value of 3.6. For the laser-assisted functionalization, the DzBr solution was diluted (1:100) to a concentration of 35 nmol/g for exposure to the nanotube.

Experimental setup for \textit{in-situ} monitoring of functionalization. A Schematic view of the laser-assisted kinetic measurements is given in Figure \ref{FIG:RM}a. The substrate was placed in a plastic container for the LAF step. The drop casted CNTs were covered with 2 ml of deionized water and a glass cover floating on the droplet to overcome the lens effect at the water surface. The measurements were done with excitation powers between 2 and 40 mW with a 5x objective resulting in a power density from 7.9 to 157 W/mm$^2$. No spectral changes were observed before Dz compound was added even at the highest laser power.

The \textit{light sensitivity} was probed by lateral Raman mapping. We mapped the defect density in the entire area around the functionalization spot. After LAF (2mW; 50x objective; 2h), a Raman mapping of the irradiated zone and its surroundings was performed with an Xplora Horiba spectrometer. The same excitation energy as for LAF (785 nm) was used, but an objective with higher magnification (x100, N.A. 0.9) was picked to obtain higher lateral resolution. At each $x$, $y$ position we acquired a Raman spectrum, containing the G and D modes. These modes were fitted with Lorentzian profiles and the intensity (integrated peak area) of the D mode was divided by the intensity of the G$^+$ mode (LO phonon) shown in Figure \ref{FIG:RM}b and \ref{FIG:Kin}d. 

We probed \textit{chiral selectivity} by comparing targeted and non-targeted species. We analyze the evolution of transition energies with resonant Raman spectroscopy. The RBMs were recorded before and after LAF (2mW; 50x objective; 2h) functionalization. A tunable excitation system was used with a Ti:Sa laser (Coherent MBR 110) providing excitations from 700 to 1000 nm. The backscattered light was collected by a 50x objective before the functionalization and with an 100x objective after the functionalization to select nanotubes from the reaction centre. A triple grating system of a T64000 Horiba spectrometer equipped with a 900 line per mm grating and a silicon charge-coupled device was used to analyse the backscattered light. Raman shift and intensity were calibrated on a sulphur reference sample. Resonant Raman maps in Figure \ref{FIG:RR}g and h were constructed from calibrated Raman spectral fits. 

To demonstrate our technique we \textit{fabricated a single tube device} and evaluated functionalization induced changes in its optical and electrical properties. Devices were prepared from commercial substrates (Active Business Company), which consist of a boron-doped silicon carrier wafer (resistance $\Omega <$ 0.005 cm) covered with 300 nm of thermal silicon oxide. The wafer was diced to 10x10 mm$^2$ and Pd electrodes with 500 nm gap size were defined by standard electron beam lithography. The toluene-based suspension containing a few-chirality semiconducting nanotubes with diameters 1- 1.2 nm, predominatly (9,7), (8,7), and (8,6) was used. Individual CNTs were deposited from solution onto multiple contact pairs by capacitive coupled ac-dielectrophoresis.\cite{Vijayaraghavan2007} The suspension was diluted to a concentration of 1 CNT/$\mu$m$^3$ solvent to deposit individual tubes, and a 20 $\mu$l droplet was placed onto the device array. A bias between 0.1 and 2 V at frequencies between 100 kHz and 1 MHz was applied between the common drain electrode and the back gate using Agilent 33250 function generator. After 5 minutes the sample was rinsed with toluene to get rid of the excess polymer and annealed for 1.5 hours at 160 °C to improve the contact conductivity. To confirm the deposition of individual CNTs transport characteristics of the devices were measured at ambient conditions in a probe station with TRIAX probes using an Agilent 4155C Semiconductor Parameter Analyzer. For current biasing, we have used for the source and the drain electrode separate source-measurement-units (SMU) and operated the Agilent 4155C in constant current mode. The drain electrode has been set as a reference for the gate voltage which has been applied by a third SMU in constant voltage mode. A Poly(methyl methacrylate) (PMMA) mask was deposited on top of device with a 50 nm window between the electrodes.

For applying our \textit{LAF method to our device}, the substrate with the devices was covered with the DzBr-solution for 45 minutes and irradiated with 785 nm at 40 mW and 5x magnification. The sample was then rinsed with water, blown dry with nitrogen, and annealed on a hotplate for 5 minutes at 130°C.

Finally, with \textit{electroluminescence spectroscopy} we studied the influence of the functionalization of the optical properties of the device: Samples were mounted on a custom made sample holder and chip contacts of up to eight devices were bonded onto palladium pads attached to this holder. \textit{In-situ} annealing at 60-70 °C was conducted via the integrated heater at a pressure below 10.6 mbar and the subsequent electroluminescence measurements were carried out without breaking vacuum. The cryostat has a 10 mm diameter optical access via a 0.5 mm thick quartz window and the emitted light was collected with a Zeiss LD-Plan Neofluar objective (40x/0.6) of a customized Zeiss Axiotech Vario microscope and focused with an off-axis parabolic mirror (Thorlabs MPD149-P01, Ag, 25.4 mm, f/4) into an Acton SP-2360 (f/3.9) imaging spectrograph (Princeton Instruments) and dispersed via a 85 G/mm, 1.35 $\mu m$ blazed grating onto an InGaAs photodiode linear array (PyLoN-IR Princeton Instruments) with 1024 pixels, sensitive from 950-1610 nm. The cryostat is positioned with sub-$\mu$m precision by a motorized xy scanning stage (8MTF, Standa) and the working distance between objective and sample surface is adjusted by a high precision objective piezo scanner (P-721 PIFOC / E-665 Piezo Amplifier, Physics Instruments), which allowed precise and stable positioning of the emitter. CNT-devices mounted in the cryostat were driven by an Agilent 4155B Semiconductor Parameter Analyzer. The noise in the spectra in Figure \ref{FIG:EL}b was reduced by using a moving average of 15 periods, after performing all fitting operations. 

\section{Results 1. Light sensitivity. Kinetics}

\begin{figure}
  \centering
  \includegraphics[width=14cm]{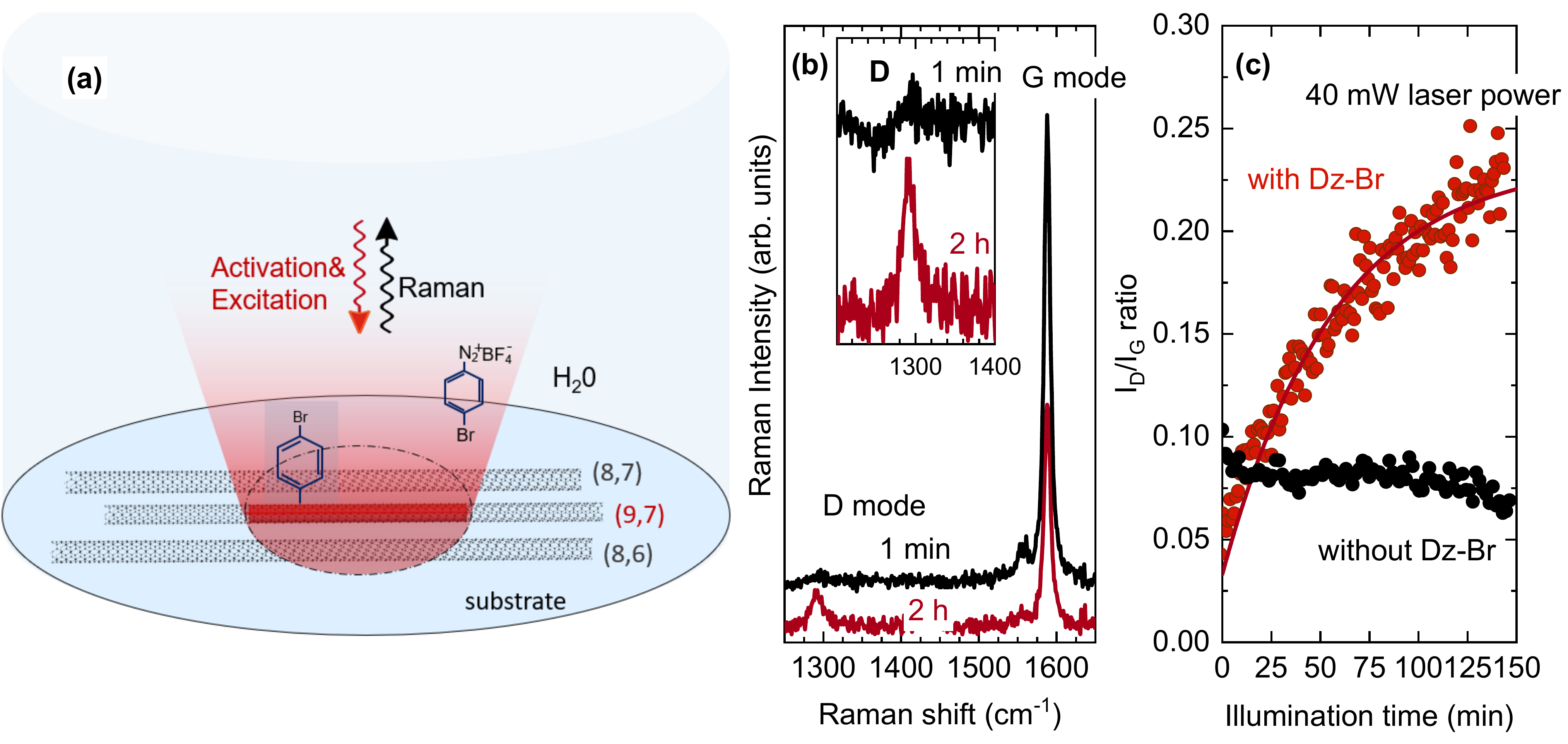}
  \caption{Experimental setup for laser assisted functionalization. (a) Experimental scheme, the nanotubes were drop casted onto the substrate and covered with the aqueous solution. The activation laser of 785nm (1.58 eV) is focused onto the CNTs and backscattered light is analyzed with a Raman spectrometer. After adding the Dz-Br to the solution the functionalization proceeds of the CNT resonant with activation light (highlighted by a red colour). (b) Examples of the Raman spectra collected during reaction (black) at start and (red) after addition of Dz-Br and 2 hours illumination. The inset shows magnified region of the D mode (c) Reference measurement: evolution of the D/G ratio as a function of 40 mW illumination time (black) without addition of Dz-Br and (red) with. }
  \label{FIG:RM}
\end{figure}

To begin we investigate the light sensitivity of the reaction and study whether the functionalization can be controlled with light. We drop casted a mixture of several semiconducting CNT chiralities onto the silicon substrate. After the CNTs adhered to the substrate, the sample was covered with water, as shown in Figure \ref{FIG:RM}a. The laser light of 1.58 eV tuned in resonance with the second excitonic transition $E_{22}$ of the (9,7) CNT, was focused onto the nanotubes. The inelastically scattered light was instantly analyzed by a Raman spectrometer. A Raman spectrum after 1 minute illumination is shown is Figure \ref{FIG:RM}b. The highest peak at 1592 cm$^{-1}$ is G$^+$ mode due to the longitudinal vibration (we further ommit $+$ symbol). The D mode near 1289 cm$^{-1}$ is proportional to the $sp^3$ defect concentration in the CNT\cite{Maultzsch2001}, and the ratio between the D and G modes scales with the defect density.\cite{Mohamed2015} We started the laser-assisted functionalization (LAF) by adding 4-bromobenzenediazonium tetrafluoroborate (Dz-Br) to the solution (35 nmol/g). In 2 hours the amount of defect centers produced by covalent functionalization increases yielding a more intense D mode, see inset in Figure \ref{FIG:RM}b. The time evolution of the intensity ratios between the D and G modes is shown in Figure \ref{FIG:RM}c and clearly reflects all the reaction details. The reference measurement is shown in Figure \ref{FIG:RM}c by black symbols. We used the same experimental conditions without adding Dz-Br. No changes occurred in the Raman spectrum, indicating that in our setup the laser light does not perturb the CNTs in the absence of the reaction agent. While during intense laser illumination one could expect CNTs oxidation \cite{Tachibana2013} or even defects healing \cite{Van2017}, these would only occur at higher laser densities\cite{Tachibana2013} and dry conditions.

\begin{figure}
  \centering
  \includegraphics[width=14cm]{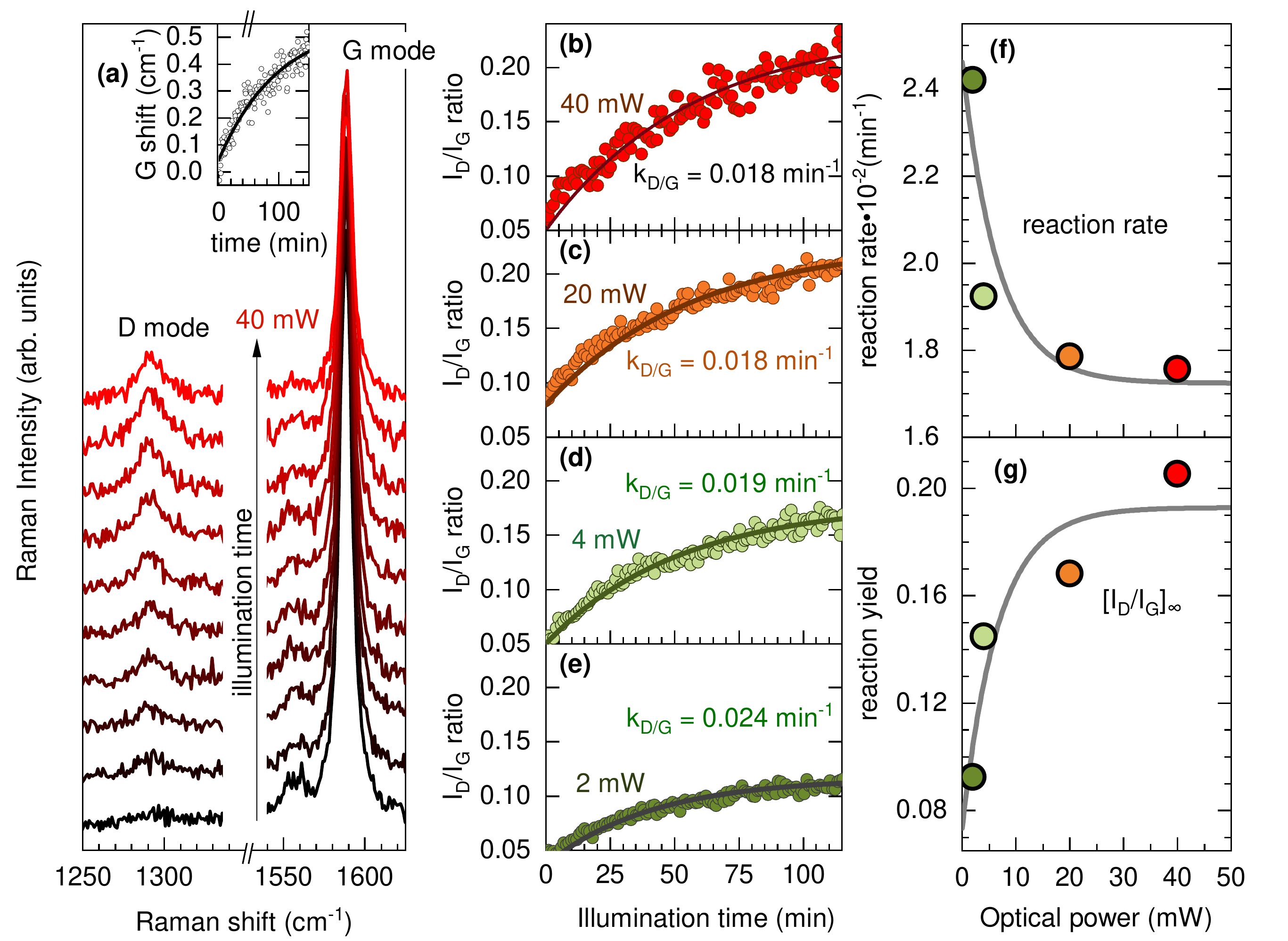}
  \caption{\textit{In-situ} control of the reaction kinetics via laser power. (a) Evolution of the Raman D and G modes during 40 mW 1.58 eV laser illumination, time increases from the bottom to the top. The inset shows time evolution of the G mode position. (b-e) The $I_D/I_G$ (integrated peak areas ratio) over illumination time for 2, 4, 20, and 40 mW laser powers. Symbols represent experimental data and line is a kinetic fit by Eq. \eqref{EQ:Kin}. (f) Reaction rate and (g) reaction yield as function of laser power. Symbols represent experimental data, and lines empirical expressions by Eq. \eqref{EQ:Rate}.
  }
  \label{FIG:Kin}
\end{figure}

Through \textit{in-situ} functionalization we can follow real-time functionalization effects. The time evolution of the CNT Raman spectra of the D and G modes during 40 mW illumination is shown in Figure \ref{FIG:Kin}a. The $I_D$ increases with time, reflecting the progress in reaction. At the same time, the G mode intensity shifts to higher energy, see inset in Figure \ref{EQ:Kin}a. The shift  of 0.5 cm$^{-1}$ is also accompanied by broadening increasing full width at half maxima by 0.7 cm$^{-1}$ as can be expected for doping induced effects.\cite{Setaro2017,Gordeev2016a} The doping is a result of additional charge carriers shared by the functional moieties. For semiconducting CNTs the doping effect are relatively weak compared to the metallic ones. Moreover it scales non-linearly with doping,\cite{Das2009,Grimm2017}. For monitoring the reaction kinetics the $\frac{I_D}{I_G}$ ratio is advantageous, since it has linear behavior with defect density and does not change upon doping \cite{Grimm2017}.

The reaction yield and product can be controlled with laser illumination power. We use the $\frac{I_D}{I_G}$ ratio for monitoring the reaction, it is shown in Figure \ref{FIG:Kin}b highest laser power used (40 mW). The D/G ratio increases from 0.07 to 0.2 within 120 min. By reducing the laser power by 20 times (2 mW) the reaction kinetics is different, see Figure \ref{FIG:Kin}e. The D/G ratio increased only up to 0.1 within 120 min and is near at the saturation point, indicating lower reaction yield but higher reaction rate, compared to the high laser power. The reaction kinetics for two intermediate steps 4 and 20 mW is shown is Figure \ref{FIG:Kin}c and d respectively, each displaying slightly different kinetics. We quantify these changes by fitting the curves with the first-order differential equation \cite{Connors1990}:
\begin{equation}
 \frac{I_D}{I_G}(t)=\frac{I_D}{I_G}_\infty(1-e^{-k_D t} )+\frac{I_D}{I_G}_0
    \label{EQ:Kin}
\end{equation}
where $\frac{I_D}{I_G}_\infty$ and $\frac{I_D}{I_G}_0$ reflect the defects density at infinite and zero reaction time respectively. $k_D$ is the reaction rate derived from the defect density. The experimental data is fitted by Eq. \eqref{EQ:Kin} and the fit is shown in Figure \ref{FIG:Kin}b-e by the solid lines. The parameters are listed in Table \ref{TAB:T1} along with the standard fit errors. The defects induced by the reaction can be associated with the reaction yield. The yield $\frac{I_D}{I_G}_\infty$ increases with photon density and can be related to the variation of activation energy, which varies non-linearly with optical power as shown in Figure \ref{FIG:Kin}g. This variation could be in theory induced by photo gating of the nanotube on the SiO$_2$/Si surface, which is chirality independent,\cite{Benfante2018,Riaz2019} we later perform the resonance Raman analysis to verify these effects. The positive effect of the light intensity onto the reaction yield is comparable with the solution based studies.\cite{Powell2016}



\begin{table}[h]
    \centering

    \caption{Laser power control parameters over the diazonium/nanotube reaction rate $k_D\pm10^{-3}$ (min$^{-1}$) and reaction yield $\frac{I_D}{I_G}_\infty\pm5\cdot10^{-4}$ from Eq. \eqref{EQ:Kin}, plotted in Figure \ref{FIG:Kin}.}
    \begin{tabular}{llllll}
    
        \toprule
    Laser power&Power density& $k_D$ & $\frac{I_D}{I_G}_\infty$ &D position & G FWHM \\
    (mW)&(W/mm$^{-2}$)&(min$^{-1}$)& &(cm$^{-1}$) & (cm$^{-1}$) \\ \midrule
    2&8&0.024&0.09&1288.8&8.6 \\
    4.&16&0.019&0.14&1289.5&8.7 \\
    20&79&0.018&0.17&1291.5&9.3 \\
    40&158&0.018&0.21&1291.7&9.4 \\
  \bottomrule
    \end{tabular}
    \label{TAB:T1}
\end{table}


The reaction rate displays a curious behavior decreasing with laser power, as shown in Figure \ref{FIG:Kin}e. We can use this dependence to better understand the reaction mechanism. In order for reaction to occur, a free aryl radical must be created first.\cite{Schmidt2009} There are two possibilities to generate a radical, (I) to generate the radical via the direct electron extraction from CNT.\cite{Dyke2004} And (II) that works through the interaction of the radical with the base,\cite{Galli1988,Gomberg1924} resulting in an  diazonium anhydride intermediate species. This species can be thermally decomposed into the aryl radical. While the (II) mechanism was generally accepted, recent experiment show that the reaction can occur without a base, when illuminated by light, pointing rather to the (I) pathway. That mechanism will require photoexcitation of the nanotube, which is more efficient near its absorption bands ($E_{ii}$ energies).  This is in line with solution-based chiral selectivity studies ascribe.\cite{Powell2016} In both (I) and (II) pathways the reaction rate must increase with higher laser power, either due to more efficient heating induced diffusion or photo-excitation\cite{Burek2019}. Instead we find that the reaction is slower for higher laser powers, see Figure \ref{FIG:Kin}f. We attribute this counter intuitive behavior to a competing parallel reaction. As commonly known, the diazonium cations decompose in water into nitrogen and phenyl when enough thermal energy is given to the system.{\cite{DeTar1956}} Higher laser energy induces more local heat development, thus accelerating the competing reaction {\cite{DeTar1956}} that effectively reduces the production of the aryl radicals. The heat is instantly carried away from the nanotube surface as no heat-induced effects are observed in the Raman spectrum during reference measurement shown in Figure \ref{FIG:RM}c. The competing reaction leads to a retardation of the functionalization due to a reduced concentration of reactant around the nanotube surface.


We find that the reaction rate $k$ and the reaction yield $C$ $\left(\frac{I_D}{I_G}_\infty \right)$ from Eq. \eqref{EQ:Kin} dependence on laser power follow the empirical expressions:

\begin{equation}
   k(P),C(P)=A_{k,C}+B_{k,C} \left( 1+e^{-k_p P} \right),
    \label{EQ:Rate}
\end{equation}
with A, B, $k_p$ as constants ($A_k=   7\cdot10^{-3} $min$^{-1}$, $A_C$ = 0.1 min$^{-1}$, $B_k$  = 0.01 min$^{-1}$, $B_C = 7\cdot10^{-2}$ min$^{-1}$, and $k_p = 1.5\cdot10^{-2}$ mW$^{-1}$). The plots of Eq. \eqref{EQ:Rate} with these parameters are shown in Figure \ref{FIG:Kin}e by full lines. They may be equally applicable for substrate supported nanotubes as well as for bulk solutions with low opacity, enabling propagation of the activation light through the entire volume. Two control parameters can be used to achieve the desired functionalization rate, the photon density and the illumination time. When a fast reaction is desired, the highest laser power should be applied. Up to now we investigated the CNTs under the laser spot, we also need to check them in the dark areas.

\begin{figure}
  \centering
  \includegraphics[width=14cm]{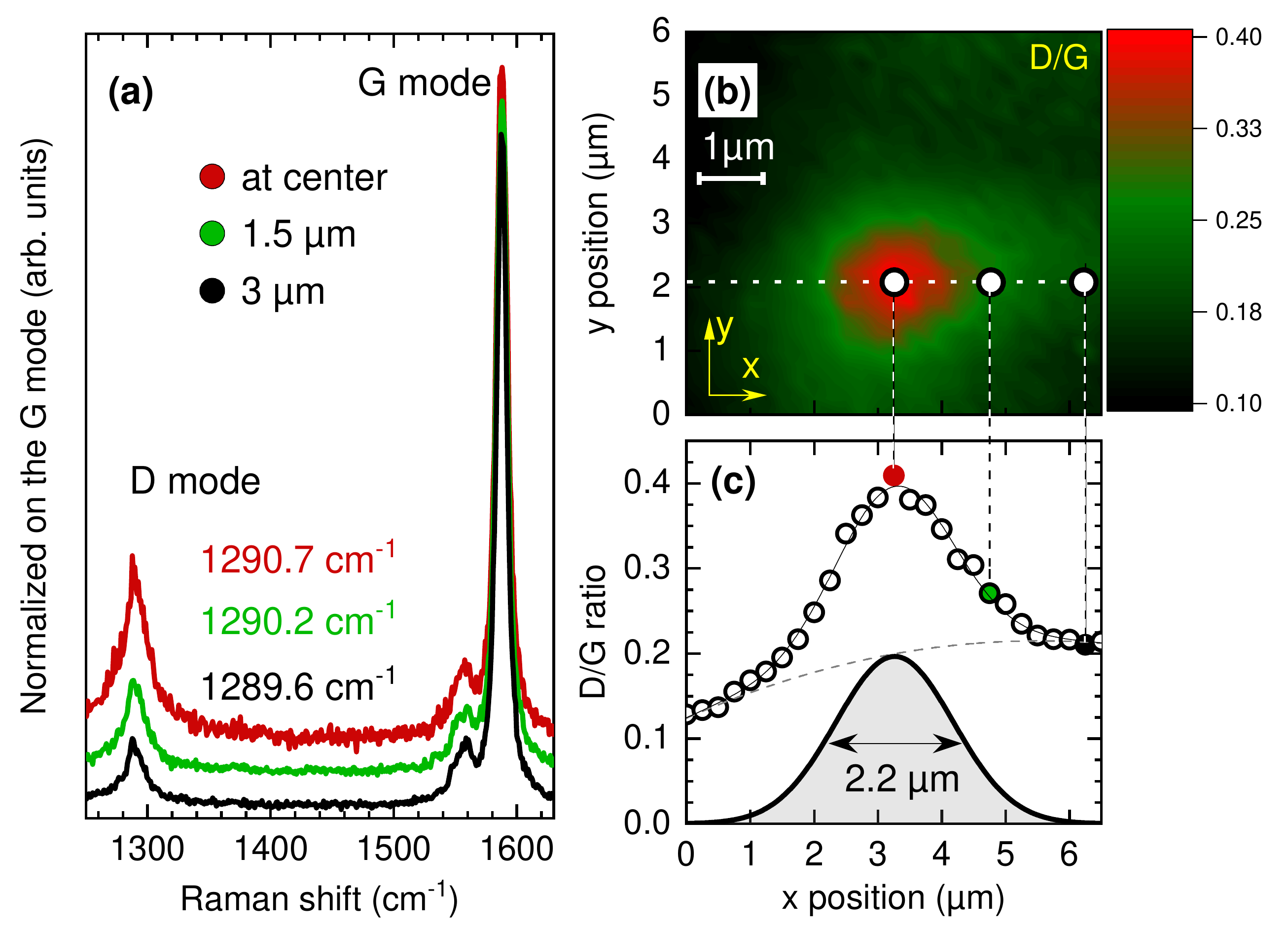}
  \caption{Light sensitivity of the reaction. (a) Raman spectra acquired at different distances from the functionalization spot center excited at 1.58 eV. (b) Lateral variation of the intensity ratio between the D and G Raman modes over the functionalization spot. Horizontal dashed line indicates where the linear profile is shown in (c). (c) The D to G profile as a function of the $x$ position, symbols represent experimental points and line is a fit by Gaussian line shape. The width of Gaussian peak corresponds to the full width at half maximum of the laser spot.}
  \label{FIG:Lat}
\end{figure}

We probed light sensitivity of the reaction by spatial Raman mapping. After the LAF, we rinsed the sample with water multiple times and dried it. We collected Raman spectra in the area around the functionalization spot (FS) using higher lateral resolution of ~600 nm, as shown in Figure \ref{FIG:Lat}. At each lateral position we evaluated the ratio between the D and G modes and assessed the functionalization degree. The D/G ratio is plotted as two dimensional Raman map in Figure \ref{FIG:Lat}b. The highest degree of functionalization (D/G = 0.4) is observed in the middle of the FS compared to the unperturbed nanotubes outside of the FS (D/G = 0.1). We estimate the functional groups concentration in the middle to be equivalent to $\sim$1:150 of molar [Dz]/[C] ratios, sufficient to produce emission from defect states.\cite{Piao2013} Figure \ref{FIG:Lat}a details the characteristic D and G$^+$ Raman modes of semiconducting nanotubes at $\sim$ 1289 and 1592 cm$^{-1}$, respectively, at three different distances from focus centre.\cite{Thomsen2007} The doping effects scale together with defect concentration, which can be best estimated by the D mode position.\cite{Das2009, Hatting2013} The D mode shifts by 1.1 cm$^{-1}$ to higher energies between the edge and the middle of FS, indicating the p-doping type.\cite{Das2009, Hatting2013}

The increase of the defect density within the functionalization spot indicates a successful laser assisted reaction. The defect density inside the focus increases from the edges to the middle as shown by the line profile of the D/G ratio in Figure \ref{FIG:Lat}c. The defect density follows the Gaussian intensity profile of the laser beam. The full width at half maximum of the profile of 2.2 $\mu$m coincides with the width of the laser spot. It suggests that the reaction yield is proportional to the photon flux density (laser power) and functionalization gradients can be generated by laser light. The above experiments were performed at the same excitation energy as activation energy (1.58 eV), thus reporting the effects in (9,7) nanotube. Now we turn to the analysis of other chiralities.

\section{Results II. Chiral selectivity. Transition energies shifts (investigation via Resonance Raman)}

\begin{figure}
  \centering
  \includegraphics[width=14cm]{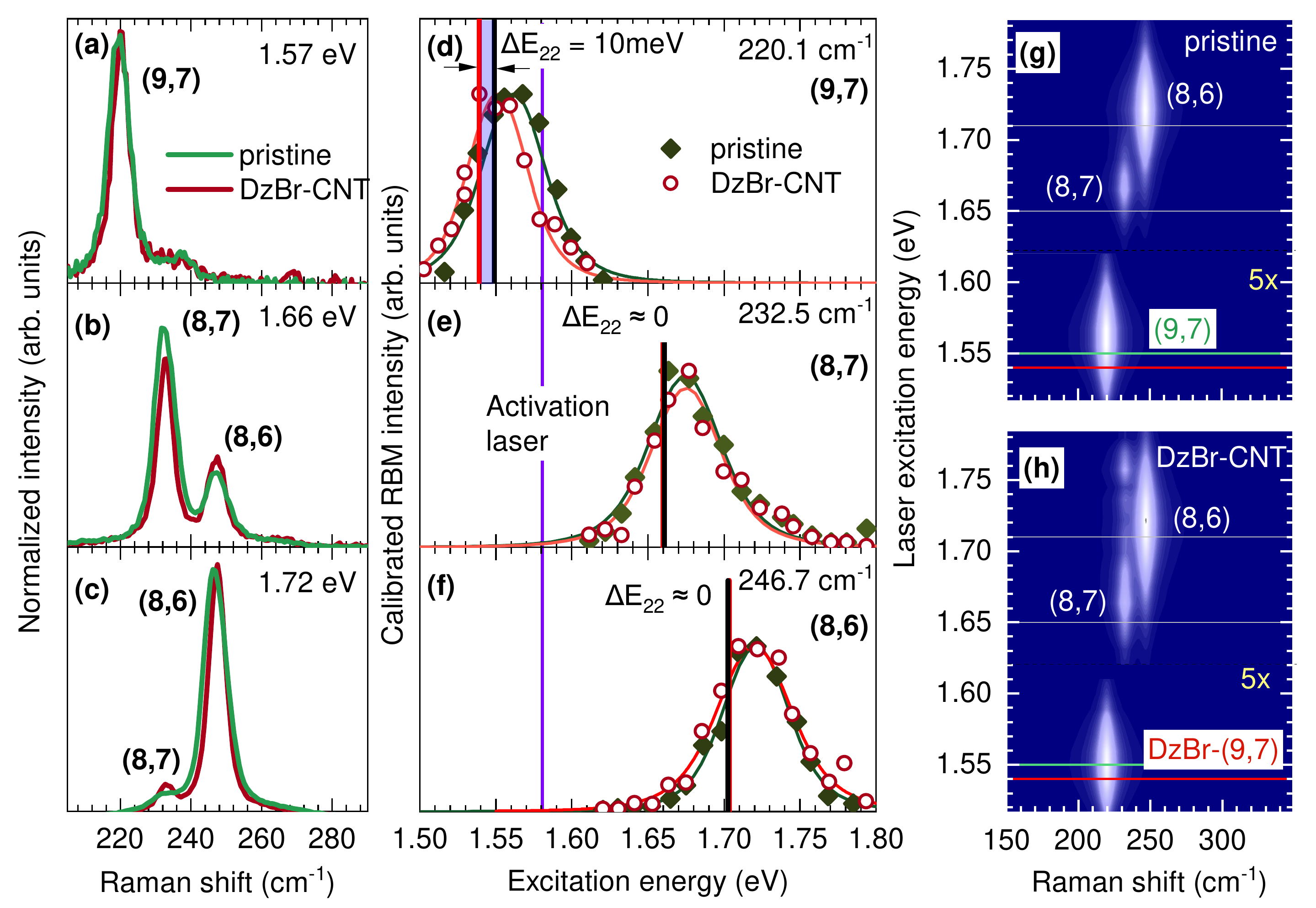}
  \caption{Chiral selectivity of the reaction. Radial breathing modes before (green) and after (red) functionalization excited at three different wavelengths (a) 1.57 eV, (b) 1.66 eV, and (c) 1.72 eV. Intensities of the RBMs as function of excitation energy (d), (e), (f) for (9,7), (8,7), and (8,6) nanotubes, respectively. The symbols represent the experimental data before (green) and after (red) functionalization. Symbols represent experimental data and lines are the fits with Eq. \eqref{EQ:RC}. Vertical colored lines mark the $E_{22}$ transition energies. The vertical purple line indicates the laser energy $E_f=1.58$ eV, used to drive the functionalization. Resonant Raman maps of (g) pristine and (h) DzBr-CNTs with indicated chiralities and transition energies marked by horizontal lines}
  \label{FIG:RR}
 \end{figure}

The G and D modes have for all nanotubes similar vibrational energies and are both excited in a broad excitation window of ~190 meV, making it impossible to distinguish between chiralities present in the sample. The energies of the radial breathing modes (RBMs) on the other hand depend on tube diameter.\cite{Maultzsch2005a} We apply resonant Raman spectroscopy to RBMs to access transition energies of each species. Raman spectra in the low energy region at three excitation energies are shown in Figure \ref{FIG:RR}a-c. The modes are index identified as the (8,6) - 247.6 cm$^{-1}$, (8,7) - 232.5 cm$^{-1}$, and (9,7) - 219.5 cm$^{-1}$.\cite{Maultzsch2005a} We measured the intensity of each mode as a function of laser excitation energies, see in Fig. \ref{FIG:RR}d-f. These Raman profiles are described by third-order perturbation theory \cite{Maultzsch2005a,Gordeev2017}:

\begin{equation}
    I_{Raman}(E_{las})=E_{las}^4{\left[ \frac{M^2_{exc-ph}M_{exc-RBM}}{
    (E_{las}-E_{22}-i\gamma)(E_{las}-\hbar\omega_{RBM}-E_{22}-i\gamma)} \right] }^2
    ,
    \label{EQ:RC}
\end{equation}

where $M_c$ is the matrix element combining exciton-photon and exciton-phonon coupling. $E_{22}$ is the  energy of the second excitonic state and $\gamma$ is the broadening factor related to the finite lifetime of the exciton. The non-resonant $E_{las}^4$ dependence in Eq. \eqref{EQ:RC} is eliminated by calibration on the Raman reference sample. The fits are shown by the solid lines in Figure \ref{FIG:RR}d-f and the parameters are listed in Table \ref{TAB:T2}. The transition energies of the pristine and functionalized (9,7) CNT (vertical lines) differ by 10 meV, whereas the transition energy of the (8,6) and (8,7) CNTs remain constant. The difference in transition energies can be also viewed on resonance Raman maps, see Figure \ref{FIG:RR}g,h. The vertical axis represents the laser excitation energy and horizontal Raman shift and the blue (red) colour represents low (high) Raman intensity. The vertical positions of the resonant peaks are different for the DzBr-(9,7) and (9,7), see horizontal lines in Figure \ref{FIG:RR}g,f, whereas the positions of the (8,6) and (8,7) do not change after functionalization.

We attribute the shift of transition energies to the dielectric effects. During covalent functionalization the functional groups either directly influence the exciton screening or indirectly change of the polymer wrapping. A denser dielectric environment leads to a red-shift of the optical transition energies.\cite{Perebeinos2004,Walsh2007,Araujo2009} The shifts of transition energies are broadly used to confirm covalent and non-covalent functionalization.\cite{Roquelet2010, Setaro2017} But the increase of the D mode (Fig. \ref{FIG:Kin}a) unambiguously indicates covalent type and the $E_{ii}$ shift may serve as an additional effect indicating functionalization. Non-targeted species (8,6) and (8,7) do not display any shift of transition energy, but might still be functionalized to some degree, certainly weaker compared to the targeted (9,7) species, see Figure \ref{FIG:RR}.  

We suggest the charge carriers photo-excitation to be responsible for the chiral selectivity mechanism. While the local thermal activation mechanism\cite{Powell2016} would yield similar effects, it is not compatible with base-free reactions\cite{Huang2020} and would might be less chirality selective, since the heat is quickly distributed within the bundle. In direct photo-excitation the laser generates excitons that partly relax into charge carriers. These charge carriers participate in the aryl radical formation\cite{Dyke2004}, yielding higher reactivity of photo excited CNTs. With this interpretation we conclude that the reaction yield is proportional to the absorption overlay with the laser energy, i.e. an additional activation occurs if $E_{f}$ is in the rage $ (E_{22}-\gamma /2,E_{22}+\gamma /2)$,  where $\gamma$ is obtained from Raman profiles, see Table \ref{TAB:T2}. The energies overlap can be directly seen by the relative energies of the RBM resonance Raman profiles and $E_{f}$ in Figure \ref{FIG:RR}d-f. Charge carriers could be also generated by the Si/SiO{$_2$} interface, but this effect appears to be less important. It should equally affect all chiralities, whereas we observe pronounced chiral selectivity in Figure \ref{FIG:RR}. The chiral selectivity of the reaction is demonstrated here between Dz-Br and (9,7) nanotubes, however, we anticipate that method to be applicable to any arbitrary (n,m) chirality and diazonium compound. Finally, we further test our apply aryl bromide functionalization on individual nanotube pre-integrated into a device.

\begin{table}[h]
    \centering
    \caption{Second transition energy $E_{22}$ for pristine nanotubes (prist) and DzBr-CNT (func) acquired by fitting the resonance Raman profiles in Figure \ref{FIG:RR}a-c, detuning is $E_{22}$-$E_f$ ($E_f=1.58$ eV) and $E_{22}$ shift = $E_{22}^{func}-E_{22}^{prist}$. The average $E_{22}$ ($\gamma$) error is $\pm$1.5 (3.5) meV.}
    \begin{tabular}{llllllll}
            \toprule
    (n,m)&RBM position & detuning & $E_{22}^{prit}$ & $E_{22}^{func}$ & $E_{22}$ shift& $\gamma_{prist}$ & $ \gamma_{func}$ \\
     & (cm$^{-1})$& (meV) & (eV)& (eV)& (meV)&(meV)&(meV) \\\midrule
    (8,6)&248&84&1.65&1.65&-0.7&32&37 \\
    (8,7)&233&136&1.71&1.71&0.8&31&39 \\
    (9,7)&220&-22&1.55&1.54&-10.5&32&28 \\

  \bottomrule
    \end{tabular}
    \label{TAB:T2}
\end{table}

\section{Results III. Device integration, electroluminescence}
\begin{figure}
  \centering
  \includegraphics[width=14cm]{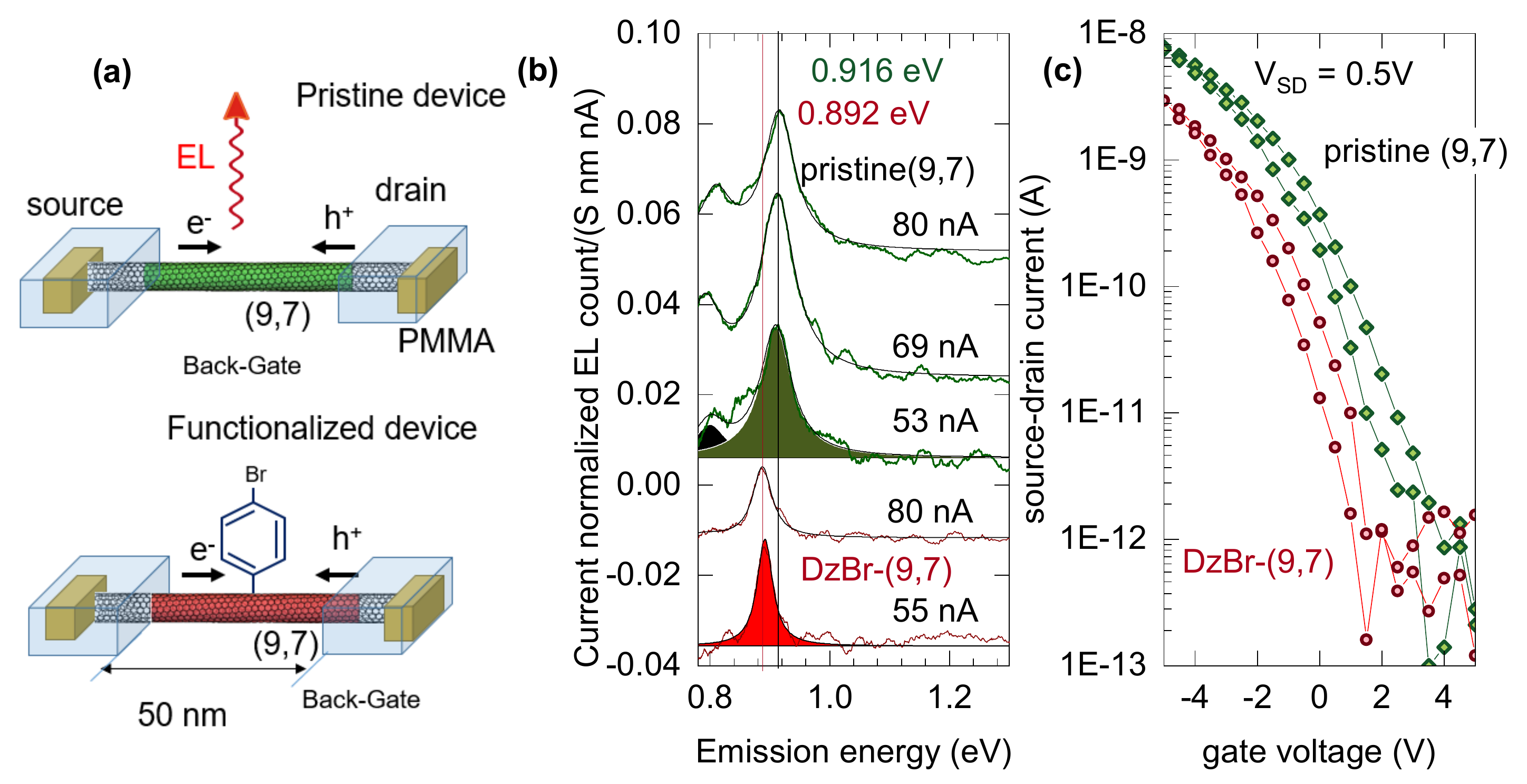}
  \caption{Aryl bromide functionalization of an electroluminescent (9,7) nanotube device. (a) Device sketch and geometry. (b) Electroluminescence spectra of (9, 7) nanotubes at different currents before (green) after functionalization (red). Peaks were fitted by a single Lorentzian (see filled peaks). (c) Transconductance curves measured before and after functionalization}
  \label{FIG:EL}
 \end{figure}
 
Now we discuss the application of the LAF method to a CNT device. A sketch of the device is shown in Figure \ref{FIG:EL}a, where a nanotube is positioned between two electrodes by dielectrophoresis, see methods section. An equal amount of holes and electrons is injected through opposite electrodes, followed by recombination in the middle and emerging photoluminescence.\cite{Marquardt2010,Gaulke2020} A spectrum of emitted light from such a device prior functionalization is shown in Figure \ref{FIG:EL}b (green trace). The peaks around 0.9 eV and 0.8 eV are associated with the emission from an $E_{11}$ exciton (electron-hole pair) \cite{Bachilo2002} and a trion (charged exciton) of the (9,7) $E_{11}$ state, as discussed by \textit{Gaulke et al.}\cite{Gaulke2020} The absence of other peaks indicates that this device most likely comprises only a single nanotube. The position of the $E_{11}$ emission varies by up to 3 meV when driven at different DC currents.

As a next step we performed the LAF directly on device. In order to ensure that aryl ring binds the nanotube in the middle, we applied a PMMA mask with a 50nm slit centred between the two electrodes, see methods. The emission of the 4-Bromophenyl functionalized device was collected for a series of currents as shown in Figure \ref{FIG:EL}b by red colour. The emission of the ground $E_{11}$ excitonic state is found at 0.92 eV, which is red-shifted by approximately 20 meV compared to the emission of the pristine device. The shift of transition energy indicates successful functionalization. A similar effect we observed in the $E_{22}$ transition by the resonance Raman spectroscopy, see Fig. \ref{FIG:RR}. On the other hand, the shift correlates with a D to G intensity ratio of at least 0.2 (Fig. \ref{FIG:Kin}a,d) and indicates that covalent bonds for brightening of the dark emission states have been formed.\cite{Piao2013} The emission of the defect state is known to be  ~180 meV below $E_{11}$ ground excitonic bright state and is expected at 0.73 eV\cite{Piao2013} and is outside of our detection range.

Aryl bromide functionalization changes the doping level of the nanotubes. This manifests in a suppression of the (9,7) trion (charged exciton) at 0.8 eV emission. However this effect is stronger than expected, additionally, the $E_{11}$ trion might be suppressed by the defect formation, as reported for two-dimensional excitons.\cite{Greben2020} Further, the Fermi level shift can be seen in the transconductance curves, shown in Figure \ref{FIG:EL}c. The p-type conductance is typical for CNTs in contact with Pd electrodes mounted onto SiO$_2$/Si surfaces due to the Schottky barrier for holes being smaller than for electrons. The positive threshold voltage for current switching indicates initial p-doping of the device which reduces to lower gate voltages after functionalization. Only from transport data we cannot rule out non-functionalization related doping origins, such as filling and depletion of trapped states. But in the Raman analysis we find similar effects where a broadening and shift of the G and D modes was observed. Thus, the functionalization of semiconducting CNTs is detectable in electrical transport data, as have been previously reported for metallic nanotubes.\cite{Wilson2016,Lee2018} The on/off currents of our device do not change much, see Figure \ref{FIG:EL}c. We conclude that the covalent functionalization is not sufficient to hinder the conductance of semiconducting CNTs, in contrast to the metallic ones\cite{Goldsmith2007}. 

The origin of the red-shift both of the second excitonic state (obtained via Raman excitation mapping) and the first excitonic state (obtained by EL spectroscopy) could be induced by the dielectric screening of the excitons.\cite{Araujo2009} The screening might be provided by the aryl rings covering the nanotube surface. An alternative physical effect providing the shift of excitonic levels can be strain,\cite{Setaro2017,Gordeev2016a} however, strain along the nanotube axis causes shifts of the second ($E_{22}$) and first states ($E_{11}$) in opposite directions,\cite{Berger2009} which is not the case in our system. Therefore dielectric screening appears to be the dominant mechanism influencing optical properties of functionalized CNTs. 
Note that the lateral control of the reaction centre was improved past the diffraction limit \cite{Huang2020} by applying polymer masks and electron beam lithography. The next step is the functionalization of devices build upon small diameter enriched semiconducting nanotubes\cite{Li2019,Graf2016}, where the defect PL is better detectable. Furthermore, the on device functionalization scheme can be of use for other $sp^2$ hybridized nanocarbons with optical transitions in the visible wavelength range such as twisted graphene layers, graphene nanoribbons, and graphene quantum dots.\cite{Denk2014,Havener2014,Wang2017}

\section{Conclusions}
A novel method that enables a local and chirality-selective covalent functionalization of nanotubes on surfaces and in devices was successfully demonstrated. Laser assisted functionalization of (9,7) nanotube with 4-bromophenyl was studied by spatial Raman mapping of defect density, which confirmed the light sensitivity of the reaction. The nanotubes inside the illumination spot were functionalized, whereas the non-illuminated areas remained unperturbed. The influence of the photon flux density on reaction kinetics was explored by \textit{in-situ} reaction monitoring. The reaction product increased and reaction rate decreased with laser power. The drop of reaction rate with the increase of photon flux reveals a competing mechanism, and could be due to Diazonium salt decomposition activated by local heat. We proposed empirical expressions for controlling the reaction kinetics applicable both to the processes on the surface and in solutions with low opacity.

To test chiral selectivity of the reaction, we compared $E_{22}$ transition energies of the pristine and functionalized nanotubes before and after functionalization. The transition energies were determined by resonant Raman spectroscopy. The $E_{22}$ of (9,7) nanotube red shifted by 10 meV in contrast with $E_{22}$ of the (8,6) and (8,7) retaining their positions. These two chiralities were out of resonance with the laser energy during functionalization. Chiral selectivity was achieved by matching the laser light energy with the optical transition of the (9.7) nanotube only.

We applied the laser-assisted aryl bromide functionalization to an electroluminescent device made of (9,7) nanotube. The reaction centre was controlled beyond the diffraction limit by a 50 nm opened PMMA slit mark. The functionalization yielded a 25 meV red-shift of the ground optical state ($E_{11}$) electroluminescence. A red shift of the first (EL) and the second (resonance Raman) transitions shows evidences of dielectric screening of the exciton wave function by molecules. The transport characteristics of the functionalized device and Raman analysis suggest a doping effect provided by the functional group. We therefore conclude that laser-assisted on device covalent functionalization opens a route towards nanoscale engineering of trapped excitonic states for optoelectronic circuits.

\textbf{Acknowledgements}
G.G. and S.R. acknowledge Focus Area NanoScale of Freie Universitaet Berlin. S. R. acknowledges support by the Deutsche Forschungsgemeinschaft under SPP 2244 and the the German Research Foundation  (DFG via SFB 658, subproject A6). We thank Oisín Garrity for language checkup. FH and RK acknowledge support by the Helmholtz Association and the Volkswagen 
Foundation.

\bibliography{bibliography}

\begin{thebibliography}{62}%
\makeatletter
\providecommand \@ifxundefined [1]{%
 \@ifx{#1\undefined}
}%
\providecommand \@ifnum [1]{%
 \ifnum #1\expandafter \@firstoftwo
 \else \expandafter \@secondoftwo
 \fi
}%
\providecommand \@ifx [1]{%
 \ifx #1\expandafter \@firstoftwo
 \else \expandafter \@secondoftwo
 \fi
}%
\providecommand \natexlab [1]{#1}%
\providecommand \enquote  [1]{``#1''}%
\providecommand \bibnamefont  [1]{#1}%
\providecommand \bibfnamefont [1]{#1}%
\providecommand \citenamefont [1]{#1}%
\providecommand \href@noop [0]{\@secondoftwo}%
\providecommand \href [0]{\begingroup \@sanitize@url \@href}%
\providecommand \@href[1]{\@@startlink{#1}\@@href}%
\providecommand \@@href[1]{\endgroup#1\@@endlink}%
\providecommand \@sanitize@url [0]{\catcode `\\12\catcode `\$12\catcode
  `\&12\catcode `\#12\catcode `\^12\catcode `\_12\catcode `\%12\relax}%
\providecommand \@@startlink[1]{}%
\providecommand \@@endlink[0]{}%
\providecommand \url  [0]{\begingroup\@sanitize@url \@url }%
\providecommand \@url [1]{\endgroup\@href {#1}{\urlprefix }}%
\providecommand \urlprefix  [0]{URL }%
\providecommand \Eprint [0]{\href }%
\providecommand \doibase [0]{https://doi.org/}%
\providecommand \selectlanguage [0]{\@gobble}%
\providecommand \bibinfo  [0]{\@secondoftwo}%
\providecommand \bibfield  [0]{\@secondoftwo}%
\providecommand \translation [1]{[#1]}%
\providecommand \BibitemOpen [0]{}%
\providecommand \bibitemStop [0]{}%
\providecommand \bibitemNoStop [0]{.\EOS\space}%
\providecommand \EOS [0]{\spacefactor3000\relax}%
\providecommand \BibitemShut  [1]{\csname bibitem#1\endcsname}%
\let\auto@bib@innerbib\@empty
\bibitem [{\citenamefont {Thomsen}\ and\ \citenamefont
  {Reich}(2007)}]{Thomsen2007}%
  \BibitemOpen
  \bibfield  {author} {\bibinfo {author} {\bibfnamefont {C.}~\bibnamefont
  {Thomsen}}\ and\ \bibinfo {author} {\bibfnamefont {S.}~\bibnamefont
  {Reich}},\ }\href@noop {} {\emph {\bibinfo {title} {Light Scattering in
  Solids IX}}},\ edited by\ \bibinfo {editor} {\bibfnamefont {C.}~\bibnamefont
  {Manuel}}\ and\ \bibinfo {editor} {\bibfnamefont {R.}~\bibnamefont {Merlin}}\
  (\bibinfo  {publisher} {Springer},\ \bibinfo {address} {Berlin Heidelberg},\
  \bibinfo {year} {2007})\ Chap.\ \bibinfo {chapter} {Raman Scattering in
  Carbon Nanotubes}, pp.\ \bibinfo {pages} {164--169}\BibitemShut {NoStop}%
\bibitem [{\citenamefont {He}\ \emph {et~al.}(2018)\citenamefont {He},
  \citenamefont {Htoon}, \citenamefont {Doorn}, \citenamefont {Pernice},
  \citenamefont {Pyatkov}, \citenamefont {Krupke}, \citenamefont {Jeantet},
  \citenamefont {Chassagneux},\ and\ \citenamefont {Voisin}}]{He2018}%
  \BibitemOpen
  \bibfield  {author} {\bibinfo {author} {\bibfnamefont {X.}~\bibnamefont
  {He}}, \bibinfo {author} {\bibfnamefont {H.}~\bibnamefont {Htoon}}, \bibinfo
  {author} {\bibfnamefont {S.~K.}\ \bibnamefont {Doorn}}, \bibinfo {author}
  {\bibfnamefont {W.~H.}\ \bibnamefont {Pernice}}, \bibinfo {author}
  {\bibfnamefont {F.}~\bibnamefont {Pyatkov}}, \bibinfo {author} {\bibfnamefont
  {R.}~\bibnamefont {Krupke}}, \bibinfo {author} {\bibfnamefont
  {A.}~\bibnamefont {Jeantet}}, \bibinfo {author} {\bibfnamefont
  {Y.}~\bibnamefont {Chassagneux}},\ and\ \bibinfo {author} {\bibfnamefont
  {C.}~\bibnamefont {Voisin}},\ }\bibfield  {title} {\bibinfo {title} {{Carbon
  nanotubes as emerging quantum-light sources}},\ }\href
  {https://doi.org/10.1038/s41563-018-0109-2} {\bibfield  {journal} {\bibinfo
  {journal} {Nature Materials}\ }\textbf {\bibinfo {volume} {17}},\ \bibinfo
  {pages} {663} (\bibinfo {year} {2018})}\BibitemShut {NoStop}%
\bibitem [{\citenamefont {Sun}\ \emph {et~al.}(2002)\citenamefont {Sun},
  \citenamefont {Fu}, \citenamefont {Lin},\ and\ \citenamefont
  {Huang}}]{Sun2002}%
  \BibitemOpen
  \bibfield  {author} {\bibinfo {author} {\bibfnamefont {Y.~P.}\ \bibnamefont
  {Sun}}, \bibinfo {author} {\bibfnamefont {K.}~\bibnamefont {Fu}}, \bibinfo
  {author} {\bibfnamefont {Y.}~\bibnamefont {Lin}},\ and\ \bibinfo {author}
  {\bibfnamefont {W.}~\bibnamefont {Huang}},\ }\bibfield  {title} {\bibinfo
  {title} {{Functionalized carbon nanotubes: Properties and applications}},\
  }\href {https://doi.org/10.1021/ar010160v} {\bibfield  {journal} {\bibinfo
  {journal} {Accounts of Chemical Research}\ }\textbf {\bibinfo {volume}
  {35}},\ \bibinfo {pages} {1096} (\bibinfo {year} {2002})}\BibitemShut
  {NoStop}%
\bibitem [{\citenamefont {Ernst}\ \emph {et~al.}(2012)\citenamefont {Ernst},
  \citenamefont {Heek}, \citenamefont {Setaro}, \citenamefont {Haag},\ and\
  \citenamefont {Reich}}]{Ernst2012}%
  \BibitemOpen
  \bibfield  {author} {\bibinfo {author} {\bibfnamefont {F.}~\bibnamefont
  {Ernst}}, \bibinfo {author} {\bibfnamefont {T.}~\bibnamefont {Heek}},
  \bibinfo {author} {\bibfnamefont {A.}~\bibnamefont {Setaro}}, \bibinfo
  {author} {\bibfnamefont {R.}~\bibnamefont {Haag}},\ and\ \bibinfo {author}
  {\bibfnamefont {S.}~\bibnamefont {Reich}},\ }\bibfield  {title} {\bibinfo
  {title} {{Energy transfer in nanotube-perylene complexes}},\ }\href
  {https://doi.org/10.1002/adfm.201200784} {\bibfield  {journal} {\bibinfo
  {journal} {Advanced Functional Materials}\ }\textbf {\bibinfo {volume}
  {22}},\ \bibinfo {pages} {3921} (\bibinfo {year} {2012})}\BibitemShut
  {NoStop}%
\bibitem [{\citenamefont {Piao}\ \emph {et~al.}(2013)\citenamefont {Piao},
  \citenamefont {Meany}, \citenamefont {Powell}, \citenamefont {Valley},
  \citenamefont {Kwon}, \citenamefont {Schatz},\ and\ \citenamefont
  {Wang}}]{Piao2013}%
  \BibitemOpen
  \bibfield  {author} {\bibinfo {author} {\bibfnamefont {Y.}~\bibnamefont
  {Piao}}, \bibinfo {author} {\bibfnamefont {B.}~\bibnamefont {Meany}},
  \bibinfo {author} {\bibfnamefont {L.~R.}\ \bibnamefont {Powell}}, \bibinfo
  {author} {\bibfnamefont {N.}~\bibnamefont {Valley}}, \bibinfo {author}
  {\bibfnamefont {H.}~\bibnamefont {Kwon}}, \bibinfo {author} {\bibfnamefont
  {G.~C.}\ \bibnamefont {Schatz}},\ and\ \bibinfo {author} {\bibfnamefont
  {Y.}~\bibnamefont {Wang}},\ }\bibfield  {title} {\bibinfo {title}
  {{Brightening of carbon nanotube photoluminescence through the incorporation
  of sp 3 defects}},\ }\href {https://doi.org/10.1038/nchem.1711} {\bibfield
  {journal} {\bibinfo  {journal} {Nature Chemistry}\ }\textbf {\bibinfo
  {volume} {5}},\ \bibinfo {pages} {840} (\bibinfo {year} {2013})},\ \Eprint
  {https://arxiv.org/abs/NIHMS150003} {arXiv:NIHMS150003} \BibitemShut
  {NoStop}%
\bibitem [{\citenamefont {He}\ \emph {et~al.}(2019)\citenamefont {He},
  \citenamefont {Sun}, \citenamefont {Gifford}, \citenamefont {Tretiak},
  \citenamefont {Piryatinski}, \citenamefont {Li}, \citenamefont {Htoon},\ and\
  \citenamefont {Doorn}}]{He2019}%
  \BibitemOpen
  \bibfield  {author} {\bibinfo {author} {\bibfnamefont {X.}~\bibnamefont
  {He}}, \bibinfo {author} {\bibfnamefont {L.}~\bibnamefont {Sun}}, \bibinfo
  {author} {\bibfnamefont {B.~J.}\ \bibnamefont {Gifford}}, \bibinfo {author}
  {\bibfnamefont {S.}~\bibnamefont {Tretiak}}, \bibinfo {author} {\bibfnamefont
  {A.}~\bibnamefont {Piryatinski}}, \bibinfo {author} {\bibfnamefont
  {X.}~\bibnamefont {Li}}, \bibinfo {author} {\bibfnamefont {H.}~\bibnamefont
  {Htoon}},\ and\ \bibinfo {author} {\bibfnamefont {S.~K.}\ \bibnamefont
  {Doorn}},\ }\bibfield  {title} {\bibinfo {title} {{Intrinsic limits of
  defect-state photoluminescence dynamics in functionalized carbon
  nanotubes}},\ }\href {https://doi.org/10.1039/c9nr02175b} {\bibfield
  {journal} {\bibinfo  {journal} {Nanoscale}\ }\textbf {\bibinfo {volume}
  {11}},\ \bibinfo {pages} {9125} (\bibinfo {year} {2019})}\BibitemShut
  {NoStop}%
\bibitem [{\citenamefont {Kwon}\ \emph {et~al.}(2016)\citenamefont {Kwon},
  \citenamefont {Furmanchuk}, \citenamefont {Kim}, \citenamefont {Meany},
  \citenamefont {Guo}, \citenamefont {Schatz},\ and\ \citenamefont
  {Wang}}]{Kwon2016}%
  \BibitemOpen
  \bibfield  {author} {\bibinfo {author} {\bibfnamefont {H.}~\bibnamefont
  {Kwon}}, \bibinfo {author} {\bibfnamefont {A.}~\bibnamefont {Furmanchuk}},
  \bibinfo {author} {\bibfnamefont {M.}~\bibnamefont {Kim}}, \bibinfo {author}
  {\bibfnamefont {B.}~\bibnamefont {Meany}}, \bibinfo {author} {\bibfnamefont
  {Y.}~\bibnamefont {Guo}}, \bibinfo {author} {\bibfnamefont {G.~C.}\
  \bibnamefont {Schatz}},\ and\ \bibinfo {author} {\bibfnamefont
  {Y.}~\bibnamefont {Wang}},\ }\bibfield  {title} {\bibinfo {title}
  {{Molecularly Tunable Fluorescent Quantum Defects}},\ }\href
  {https://doi.org/10.1021/jacs.6b03618} {\bibfield  {journal} {\bibinfo
  {journal} {Journal of the American Chemical Society}\ }\textbf {\bibinfo
  {volume} {138}},\ \bibinfo {pages} {6878} (\bibinfo {year}
  {2016})}\BibitemShut {NoStop}%
\bibitem [{\citenamefont {Ma}\ \emph {et~al.}(2014)\citenamefont {Ma},
  \citenamefont {Adamska}, \citenamefont {Yamaguchi}, \citenamefont {Yalcin},
  \citenamefont {Tretiak}, \citenamefont {Doorn},\ and\ \citenamefont
  {Htoon}}]{Ma2014}%
  \BibitemOpen
  \bibfield  {author} {\bibinfo {author} {\bibfnamefont {X.}~\bibnamefont
  {Ma}}, \bibinfo {author} {\bibfnamefont {L.}~\bibnamefont {Adamska}},
  \bibinfo {author} {\bibfnamefont {H.}~\bibnamefont {Yamaguchi}}, \bibinfo
  {author} {\bibfnamefont {S.~E.}\ \bibnamefont {Yalcin}}, \bibinfo {author}
  {\bibfnamefont {S.}~\bibnamefont {Tretiak}}, \bibinfo {author} {\bibfnamefont
  {S.~K.}\ \bibnamefont {Doorn}},\ and\ \bibinfo {author} {\bibfnamefont
  {H.}~\bibnamefont {Htoon}},\ }\bibfield  {title} {\bibinfo {title}
  {{Electronic structure and chemical nature of oxygen dopant states in carbon
  nanotubes}},\ }\href {https://doi.org/10.1021/nn504553y} {\bibfield
  {journal} {\bibinfo  {journal} {ACS Nano}\ }\textbf {\bibinfo {volume} {8}},\
  \bibinfo {pages} {10782} (\bibinfo {year} {2014})}\BibitemShut {NoStop}%
\bibitem [{\citenamefont {Luo}\ \emph {et~al.}(2017)\citenamefont {Luo},
  \citenamefont {Ahmadi}, \citenamefont {Shayan}, \citenamefont {Ma},
  \citenamefont {Mistry}, \citenamefont {Zhang}, \citenamefont {Hone},
  \citenamefont {Blackburn},\ and\ \citenamefont {Strauf}}]{Luo2017}%
  \BibitemOpen
  \bibfield  {author} {\bibinfo {author} {\bibfnamefont {Y.}~\bibnamefont
  {Luo}}, \bibinfo {author} {\bibfnamefont {E.~D.}\ \bibnamefont {Ahmadi}},
  \bibinfo {author} {\bibfnamefont {K.}~\bibnamefont {Shayan}}, \bibinfo
  {author} {\bibfnamefont {Y.}~\bibnamefont {Ma}}, \bibinfo {author}
  {\bibfnamefont {K.~S.}\ \bibnamefont {Mistry}}, \bibinfo {author}
  {\bibfnamefont {C.}~\bibnamefont {Zhang}}, \bibinfo {author} {\bibfnamefont
  {J.}~\bibnamefont {Hone}}, \bibinfo {author} {\bibfnamefont {J.~L.}\
  \bibnamefont {Blackburn}},\ and\ \bibinfo {author} {\bibfnamefont
  {S.}~\bibnamefont {Strauf}},\ }\bibfield  {title} {\bibinfo {title}
  {{Purcell-enhanced quantum yield from carbon nanotube excitons coupled to
  plasmonic nanocavities}},\ }\href
  {https://doi.org/10.1038/s41467-017-01777-w} {\bibfield  {journal} {\bibinfo
  {journal} {Nature Communications}\ }\textbf {\bibinfo {volume} {8}},\
  \bibinfo {pages} {1413} (\bibinfo {year} {2017})}\BibitemShut {NoStop}%
\bibitem [{\citenamefont {Mueller}\ \emph {et~al.}(2010)\citenamefont
  {Mueller}, \citenamefont {Kinoshita}, \citenamefont {Steiner}, \citenamefont
  {Perebeinos}, \citenamefont {Bol}, \citenamefont {Farmer},\ and\
  \citenamefont {Avouris}}]{Mueller2010}%
  \BibitemOpen
  \bibfield  {author} {\bibinfo {author} {\bibfnamefont {T.}~\bibnamefont
  {Mueller}}, \bibinfo {author} {\bibfnamefont {M.}~\bibnamefont {Kinoshita}},
  \bibinfo {author} {\bibfnamefont {M.}~\bibnamefont {Steiner}}, \bibinfo
  {author} {\bibfnamefont {V.}~\bibnamefont {Perebeinos}}, \bibinfo {author}
  {\bibfnamefont {A.~A.}\ \bibnamefont {Bol}}, \bibinfo {author} {\bibfnamefont
  {D.~B.}\ \bibnamefont {Farmer}},\ and\ \bibinfo {author} {\bibfnamefont
  {P.}~\bibnamefont {Avouris}},\ }\bibfield  {title} {\bibinfo {title}
  {{Efficient narrow-band light emission from a single carbon nanotube p-n
  diode}},\ }\href {https://doi.org/10.1038/nnano.2009.319} {\bibfield
  {journal} {\bibinfo  {journal} {Nature Nanotechnology}\ }\textbf {\bibinfo
  {volume} {5}},\ \bibinfo {pages} {27} (\bibinfo {year} {2010})}\BibitemShut
  {NoStop}%
\bibitem [{\citenamefont {Khasminskaya}\ \emph {et~al.}(2016)\citenamefont
  {Khasminskaya}, \citenamefont {Pyatkov}, \citenamefont {S{\l}owik},
  \citenamefont {Ferrari}, \citenamefont {Kahl}, \citenamefont {Kovalyuk},
  \citenamefont {Rath}, \citenamefont {Vetter}, \citenamefont {Hennrich},
  \citenamefont {Kappes}, \citenamefont {Gol'tsman}, \citenamefont {Korneev},
  \citenamefont {Rockstuhl}, \citenamefont {Krupke},\ and\ \citenamefont
  {Pernice}}]{Khasminskaya2016}%
  \BibitemOpen
  \bibfield  {author} {\bibinfo {author} {\bibfnamefont {S.}~\bibnamefont
  {Khasminskaya}}, \bibinfo {author} {\bibfnamefont {F.}~\bibnamefont
  {Pyatkov}}, \bibinfo {author} {\bibfnamefont {K.}~\bibnamefont {S{\l}owik}},
  \bibinfo {author} {\bibfnamefont {S.}~\bibnamefont {Ferrari}}, \bibinfo
  {author} {\bibfnamefont {O.}~\bibnamefont {Kahl}}, \bibinfo {author}
  {\bibfnamefont {V.}~\bibnamefont {Kovalyuk}}, \bibinfo {author}
  {\bibfnamefont {P.}~\bibnamefont {Rath}}, \bibinfo {author} {\bibfnamefont
  {A.}~\bibnamefont {Vetter}}, \bibinfo {author} {\bibfnamefont
  {F.}~\bibnamefont {Hennrich}}, \bibinfo {author} {\bibfnamefont {M.~M.}\
  \bibnamefont {Kappes}}, \bibinfo {author} {\bibfnamefont {G.}~\bibnamefont
  {Gol'tsman}}, \bibinfo {author} {\bibfnamefont {A.}~\bibnamefont {Korneev}},
  \bibinfo {author} {\bibfnamefont {C.}~\bibnamefont {Rockstuhl}}, \bibinfo
  {author} {\bibfnamefont {R.}~\bibnamefont {Krupke}},\ and\ \bibinfo {author}
  {\bibfnamefont {W.~H.}\ \bibnamefont {Pernice}},\ }\bibfield  {title}
  {\bibinfo {title} {{Fully integrated quantum photonic circuit with an
  electrically driven light source}},\ }\href
  {https://doi.org/10.1038/nphoton.2016.178} {\bibfield  {journal} {\bibinfo
  {journal} {Nature Photonics}\ }\textbf {\bibinfo {volume} {10}},\ \bibinfo
  {pages} {727} (\bibinfo {year} {2016})}\BibitemShut {NoStop}%
\bibitem [{\citenamefont {Strano}\ \emph {et~al.}(2003)\citenamefont {Strano},
  \citenamefont {Dyke}, \citenamefont {Usrey}, \citenamefont {Barone},
  \citenamefont {Allen}, \citenamefont {Shan}, \citenamefont {Kittrell},
  \citenamefont {Hauge}, \citenamefont {Tour},\ and\ \citenamefont
  {Smalley}}]{Strano2003}%
  \BibitemOpen
  \bibfield  {author} {\bibinfo {author} {\bibfnamefont {M.~S.}\ \bibnamefont
  {Strano}}, \bibinfo {author} {\bibfnamefont {C.~A.}\ \bibnamefont {Dyke}},
  \bibinfo {author} {\bibfnamefont {M.~L.}\ \bibnamefont {Usrey}}, \bibinfo
  {author} {\bibfnamefont {P.~W.}\ \bibnamefont {Barone}}, \bibinfo {author}
  {\bibfnamefont {M.~J.}\ \bibnamefont {Allen}}, \bibinfo {author}
  {\bibfnamefont {H.}~\bibnamefont {Shan}}, \bibinfo {author} {\bibfnamefont
  {C.}~\bibnamefont {Kittrell}}, \bibinfo {author} {\bibfnamefont {R.~H.}\
  \bibnamefont {Hauge}}, \bibinfo {author} {\bibfnamefont {J.~M.}\ \bibnamefont
  {Tour}},\ and\ \bibinfo {author} {\bibfnamefont {R.~E.}\ \bibnamefont
  {Smalley}},\ }\bibfield  {title} {\bibinfo {title} {{Electronic structure
  control of single-waited carbon nanotube functionalization}},\ }\href
  {https://doi.org/10.1126/science.1087691} {\bibfield  {journal} {\bibinfo
  {journal} {Science}\ }\textbf {\bibinfo {volume} {301}},\ \bibinfo {pages}
  {1519} (\bibinfo {year} {2003})}\BibitemShut {NoStop}%
\bibitem [{\citenamefont {Usrey}\ \emph {et~al.}(2005)\citenamefont {Usrey},
  \citenamefont {Lippmann},\ and\ \citenamefont {Strano}}]{Usrey2005}%
  \BibitemOpen
  \bibfield  {author} {\bibinfo {author} {\bibfnamefont {M.~L.}\ \bibnamefont
  {Usrey}}, \bibinfo {author} {\bibfnamefont {E.~S.}\ \bibnamefont
  {Lippmann}},\ and\ \bibinfo {author} {\bibfnamefont {M.~S.}\ \bibnamefont
  {Strano}},\ }\bibfield  {title} {\bibinfo {title} {{Evidence for a two-step
  mechanism in electronically selective single-walled carbon nanotube
  reactions}},\ }\href {https://doi.org/10.1021/ja0537530} {\bibfield
  {journal} {\bibinfo  {journal} {Journal of the American Chemical Society}\
  }\textbf {\bibinfo {volume} {127}},\ \bibinfo {pages} {16129} (\bibinfo
  {year} {2005})}\BibitemShut {NoStop}%
\bibitem [{\citenamefont {Mohamed}\ \emph {et~al.}(2015)\citenamefont
  {Mohamed}, \citenamefont {Salmi}, \citenamefont {Dahoumane}, \citenamefont
  {Mekki}, \citenamefont {Carbonnier},\ and\ \citenamefont
  {Chehimi}}]{Mohamed2015}%
  \BibitemOpen
  \bibfield  {author} {\bibinfo {author} {\bibfnamefont {A.~A.}\ \bibnamefont
  {Mohamed}}, \bibinfo {author} {\bibfnamefont {Z.}~\bibnamefont {Salmi}},
  \bibinfo {author} {\bibfnamefont {S.~A.}\ \bibnamefont {Dahoumane}}, \bibinfo
  {author} {\bibfnamefont {A.}~\bibnamefont {Mekki}}, \bibinfo {author}
  {\bibfnamefont {B.}~\bibnamefont {Carbonnier}},\ and\ \bibinfo {author}
  {\bibfnamefont {M.~M.}\ \bibnamefont {Chehimi}},\ }\bibfield  {title}
  {\bibinfo {title} {{Functionalization of nanomaterials with aryldiazonium
  salts}},\ }\href {https://doi.org/10.1016/j.cis.2015.07.011} {\bibfield
  {journal} {\bibinfo  {journal} {Advances in Colloid and Interface Science}\
  }\textbf {\bibinfo {volume} {225}},\ \bibinfo {pages} {16} (\bibinfo {year}
  {2015})}\BibitemShut {NoStop}%
\bibitem [{\citenamefont {Saha}\ \emph {et~al.}(2018)\citenamefont {Saha},
  \citenamefont {Gifford}, \citenamefont {He}, \citenamefont {Ao},
  \citenamefont {Zheng}, \citenamefont {Kataura}, \citenamefont {Htoon},
  \citenamefont {Kilina}, \citenamefont {Tretiak},\ and\ \citenamefont
  {Doorn}}]{Saha2018}%
  \BibitemOpen
  \bibfield  {author} {\bibinfo {author} {\bibfnamefont {A.}~\bibnamefont
  {Saha}}, \bibinfo {author} {\bibfnamefont {B.~J.}\ \bibnamefont {Gifford}},
  \bibinfo {author} {\bibfnamefont {X.}~\bibnamefont {He}}, \bibinfo {author}
  {\bibfnamefont {G.}~\bibnamefont {Ao}}, \bibinfo {author} {\bibfnamefont
  {M.}~\bibnamefont {Zheng}}, \bibinfo {author} {\bibfnamefont
  {H.}~\bibnamefont {Kataura}}, \bibinfo {author} {\bibfnamefont
  {H.}~\bibnamefont {Htoon}}, \bibinfo {author} {\bibfnamefont
  {S.}~\bibnamefont {Kilina}}, \bibinfo {author} {\bibfnamefont
  {S.}~\bibnamefont {Tretiak}},\ and\ \bibinfo {author} {\bibfnamefont {S.~K.}\
  \bibnamefont {Doorn}},\ }\bibfield  {title} {\bibinfo {title} {{Narrow-band
  single-photon emission through selective aryl functionalization of zigzag
  carbon nanotubes}},\ }\href {https://doi.org/10.1038/s41557-018-0126-4}
  {\bibfield  {journal} {\bibinfo  {journal} {Nature Chemistry}\ }\textbf
  {\bibinfo {volume} {10}},\ \bibinfo {pages} {1089} (\bibinfo {year}
  {2018})}\BibitemShut {NoStop}%
\bibitem [{\citenamefont {Powell}\ \emph {et~al.}(2016)\citenamefont {Powell},
  \citenamefont {Piao},\ and\ \citenamefont {Wang}}]{Powell2016}%
  \BibitemOpen
  \bibfield  {author} {\bibinfo {author} {\bibfnamefont {L.~R.}\ \bibnamefont
  {Powell}}, \bibinfo {author} {\bibfnamefont {Y.}~\bibnamefont {Piao}},\ and\
  \bibinfo {author} {\bibfnamefont {Y.}~\bibnamefont {Wang}},\ }\bibfield
  {title} {\bibinfo {title} {{Optical Excitation of Carbon Nanotubes Drives
  Localized Diazonium Reactions}},\ }\href
  {https://doi.org/10.1021/acs.jpclett.6b01771} {\bibfield  {journal} {\bibinfo
   {journal} {Journal of Physical Chemistry Letters}\ }\textbf {\bibinfo
  {volume} {7}},\ \bibinfo {pages} {3690} (\bibinfo {year} {2016})}\BibitemShut
  {NoStop}%
\bibitem [{\citenamefont {Powell}\ \emph {et~al.}(2017)\citenamefont {Powell},
  \citenamefont {Kim},\ and\ \citenamefont {Wang}}]{Powell2017}%
  \BibitemOpen
  \bibfield  {author} {\bibinfo {author} {\bibfnamefont {L.~R.}\ \bibnamefont
  {Powell}}, \bibinfo {author} {\bibfnamefont {M.}~\bibnamefont {Kim}},\ and\
  \bibinfo {author} {\bibfnamefont {Y.}~\bibnamefont {Wang}},\ }\bibfield
  {title} {\bibinfo {title} {{Chirality-Selective Functionalization of
  Semiconducting Carbon Nanotubes with a Reactivity-Switchable Molecule}},\
  }\href {https://doi.org/10.1021/jacs.7b05906} {\bibfield  {journal} {\bibinfo
   {journal} {Journal of the American Chemical Society}\ }\textbf {\bibinfo
  {volume} {139}},\ \bibinfo {pages} {12533} (\bibinfo {year}
  {2017})}\BibitemShut {NoStop}%
\bibitem [{\citenamefont {Kim}\ \emph {et~al.}(2007)\citenamefont {Kim},
  \citenamefont {Usrey},\ and\ \citenamefont {Strano}}]{Kim2007}%
  \BibitemOpen
  \bibfield  {author} {\bibinfo {author} {\bibfnamefont {W.~J.}\ \bibnamefont
  {Kim}}, \bibinfo {author} {\bibfnamefont {M.~L.}\ \bibnamefont {Usrey}},\
  and\ \bibinfo {author} {\bibfnamefont {M.~S.}\ \bibnamefont {Strano}},\
  }\bibfield  {title} {\bibinfo {title} {{Selective functionalization and free
  solution electrophoresis of single-walled carbon nanotubes: Separate
  enrichment of metallic and semiconducting SWNT}},\ }\href
  {https://doi.org/10.1021/cm061862n} {\bibfield  {journal} {\bibinfo
  {journal} {Chemistry of Materials}\ }\textbf {\bibinfo {volume} {19}},\
  \bibinfo {pages} {1571} (\bibinfo {year} {2007})}\BibitemShut {NoStop}%
\bibitem [{\citenamefont {Berger}\ \emph {et~al.}(2019)\citenamefont {Berger},
  \citenamefont {L{\"{u}}ttgens}, \citenamefont {Nowack}, \citenamefont
  {Kutsch}, \citenamefont {Lindenthal}, \citenamefont {Kistner}, \citenamefont
  {M{\"{u}}ller}, \citenamefont {Bongartz}, \citenamefont {Lumsargis},
  \citenamefont {Zakharko},\ and\ \citenamefont {Zaumseil}}]{Berger2019}%
  \BibitemOpen
  \bibfield  {author} {\bibinfo {author} {\bibfnamefont {F.~J.}\ \bibnamefont
  {Berger}}, \bibinfo {author} {\bibfnamefont {J.}~\bibnamefont
  {L{\"{u}}ttgens}}, \bibinfo {author} {\bibfnamefont {T.}~\bibnamefont
  {Nowack}}, \bibinfo {author} {\bibfnamefont {T.}~\bibnamefont {Kutsch}},
  \bibinfo {author} {\bibfnamefont {S.}~\bibnamefont {Lindenthal}}, \bibinfo
  {author} {\bibfnamefont {L.}~\bibnamefont {Kistner}}, \bibinfo {author}
  {\bibfnamefont {C.~C.}\ \bibnamefont {M{\"{u}}ller}}, \bibinfo {author}
  {\bibfnamefont {L.~M.}\ \bibnamefont {Bongartz}}, \bibinfo {author}
  {\bibfnamefont {V.~A.}\ \bibnamefont {Lumsargis}}, \bibinfo {author}
  {\bibfnamefont {Y.}~\bibnamefont {Zakharko}},\ and\ \bibinfo {author}
  {\bibfnamefont {J.}~\bibnamefont {Zaumseil}},\ }\bibfield  {title} {\bibinfo
  {title} {{Brightening of Long, Polymer-Wrapped Carbon Nanotubes by sp3
  Functionalization in Organic Solvents}},\ }\href
  {https://doi.org/10.1021/acsnano.9b03792} {\bibfield  {journal} {\bibinfo
  {journal} {ACS Nano}\ }\textbf {\bibinfo {volume} {13}},\ \bibinfo {pages}
  {9259} (\bibinfo {year} {2019})}\BibitemShut {NoStop}%
\bibitem [{\citenamefont {Huang}\ \emph {et~al.}(2020)\citenamefont {Huang},
  \citenamefont {Powell}, \citenamefont {Wu}, \citenamefont {Kim},
  \citenamefont {Qu}, \citenamefont {Wang}, \citenamefont {Fortner},
  \citenamefont {Xu}, \citenamefont {Ng},\ and\ \citenamefont
  {Wang}}]{Huang2020}%
  \BibitemOpen
  \bibfield  {author} {\bibinfo {author} {\bibfnamefont {Z.}~\bibnamefont
  {Huang}}, \bibinfo {author} {\bibfnamefont {L.~R.}\ \bibnamefont {Powell}},
  \bibinfo {author} {\bibfnamefont {X.}~\bibnamefont {Wu}}, \bibinfo {author}
  {\bibfnamefont {M.}~\bibnamefont {Kim}}, \bibinfo {author} {\bibfnamefont
  {H.}~\bibnamefont {Qu}}, \bibinfo {author} {\bibfnamefont {P.}~\bibnamefont
  {Wang}}, \bibinfo {author} {\bibfnamefont {J.~L.}\ \bibnamefont {Fortner}},
  \bibinfo {author} {\bibfnamefont {B.}~\bibnamefont {Xu}}, \bibinfo {author}
  {\bibfnamefont {A.~L.}\ \bibnamefont {Ng}},\ and\ \bibinfo {author}
  {\bibfnamefont {Y.~H.}\ \bibnamefont {Wang}},\ }\bibfield  {title} {\bibinfo
  {title} {{Photolithographic Patterning of Organic Color-Centers}},\ }\href
  {https://doi.org/10.1002/adma.201906517} {\bibfield  {journal} {\bibinfo
  {journal} {Advanced Materials}\ }\textbf {\bibinfo {volume} {32}},\ \bibinfo
  {pages} {1906517} (\bibinfo {year} {2020})}\BibitemShut {NoStop}%
\bibitem [{\citenamefont {Maultzsch}\ \emph {et~al.}(2001)\citenamefont
  {Maultzsch}, \citenamefont {Reich},\ and\ \citenamefont
  {Thomsen}}]{Maultzsch2001}%
  \BibitemOpen
  \bibfield  {author} {\bibinfo {author} {\bibfnamefont {J.}~\bibnamefont
  {Maultzsch}}, \bibinfo {author} {\bibfnamefont {S.}~\bibnamefont {Reich}},\
  and\ \bibinfo {author} {\bibfnamefont {C.}~\bibnamefont {Thomsen}},\
  }\bibfield  {title} {\bibinfo {title} {{Chirality-selective Raman scattering
  of the D mode in carbon nanotubes}},\ }\href
  {https://doi.org/10.1103/PhysRevB.64.121407} {\bibfield  {journal} {\bibinfo
  {journal} {Physical Review B}\ }\textbf {\bibinfo {volume} {64}},\ \bibinfo
  {pages} {121407} (\bibinfo {year} {2001})}\BibitemShut {NoStop}%
\bibitem [{\citenamefont {Thomsen}\ and\ \citenamefont
  {Reich}(2000)}]{Thomsen2000}%
  \BibitemOpen
  \bibfield  {author} {\bibinfo {author} {\bibfnamefont {C.}~\bibnamefont
  {Thomsen}}\ and\ \bibinfo {author} {\bibfnamefont {S.}~\bibnamefont
  {Reich}},\ }\bibfield  {title} {\bibinfo {title} {{Double resonant raman
  scattering in graphite}},\ }\href
  {https://doi.org/10.1103/PhysRevLett.85.5214} {\bibfield  {journal} {\bibinfo
   {journal} {Physical Review Letters}\ }\textbf {\bibinfo {volume} {85}},\
  \bibinfo {pages} {5214} (\bibinfo {year} {2000})},\ \Eprint
  {https://arxiv.org/abs/1001.4690} {arXiv:1001.4690} \BibitemShut {NoStop}%
\bibitem [{\citenamefont {Can{\c{c}}ado}\ \emph {et~al.}(2011)\citenamefont
  {Can{\c{c}}ado}, \citenamefont {Jorio}, \citenamefont {Ferreira},
  \citenamefont {Stavale}, \citenamefont {Achete}, \citenamefont {Capaz},
  \citenamefont {Moutinho}, \citenamefont {Lombardo}, \citenamefont {Kulmala},\
  and\ \citenamefont {Ferrari}}]{Cancado2011}%
  \BibitemOpen
  \bibfield  {author} {\bibinfo {author} {\bibfnamefont {L.~G.}\ \bibnamefont
  {Can{\c{c}}ado}}, \bibinfo {author} {\bibfnamefont {A.}~\bibnamefont
  {Jorio}}, \bibinfo {author} {\bibfnamefont {E.~H.}\ \bibnamefont {Ferreira}},
  \bibinfo {author} {\bibfnamefont {F.}~\bibnamefont {Stavale}}, \bibinfo
  {author} {\bibfnamefont {C.~A.}\ \bibnamefont {Achete}}, \bibinfo {author}
  {\bibfnamefont {R.~B.}\ \bibnamefont {Capaz}}, \bibinfo {author}
  {\bibfnamefont {M.~V.}\ \bibnamefont {Moutinho}}, \bibinfo {author}
  {\bibfnamefont {A.}~\bibnamefont {Lombardo}}, \bibinfo {author}
  {\bibfnamefont {T.~S.}\ \bibnamefont {Kulmala}},\ and\ \bibinfo {author}
  {\bibfnamefont {A.~C.}\ \bibnamefont {Ferrari}},\ }\bibfield  {title}
  {\bibinfo {title} {{Quantifying defects in graphene via Raman spectroscopy at
  different excitation energies}},\ }\href {https://doi.org/10.1021/nl201432g}
  {\bibfield  {journal} {\bibinfo  {journal} {Nano Letters}\ }\textbf {\bibinfo
  {volume} {11}},\ \bibinfo {pages} {3190} (\bibinfo {year} {2011})},\ \Eprint
  {https://arxiv.org/abs/1105.0175} {arXiv:1105.0175} \BibitemShut {NoStop}%
\bibitem [{\citenamefont {Can{\c{c}}ado}\ \emph {et~al.}(2017)\citenamefont
  {Can{\c{c}}ado}, \citenamefont {{Da Silva}}, \citenamefont {{Martins
  Ferreira}}, \citenamefont {Hof}, \citenamefont {Kampioti}, \citenamefont
  {Huang}, \citenamefont {P{\'{e}}nicaud}, \citenamefont {Achete},
  \citenamefont {Capaz},\ and\ \citenamefont {Jorio}}]{Cancado2017}%
  \BibitemOpen
  \bibfield  {author} {\bibinfo {author} {\bibfnamefont {L.~G.}\ \bibnamefont
  {Can{\c{c}}ado}}, \bibinfo {author} {\bibfnamefont {M.~G.}\ \bibnamefont {{Da
  Silva}}}, \bibinfo {author} {\bibfnamefont {E.~H.}\ \bibnamefont {{Martins
  Ferreira}}}, \bibinfo {author} {\bibfnamefont {F.}~\bibnamefont {Hof}},
  \bibinfo {author} {\bibfnamefont {K.}~\bibnamefont {Kampioti}}, \bibinfo
  {author} {\bibfnamefont {K.}~\bibnamefont {Huang}}, \bibinfo {author}
  {\bibfnamefont {A.}~\bibnamefont {P{\'{e}}nicaud}}, \bibinfo {author}
  {\bibfnamefont {C.~A.}\ \bibnamefont {Achete}}, \bibinfo {author}
  {\bibfnamefont {R.~B.}\ \bibnamefont {Capaz}},\ and\ \bibinfo {author}
  {\bibfnamefont {A.}~\bibnamefont {Jorio}},\ }\bibfield  {title} {\bibinfo
  {title} {{Disentangling contributions of point and line defects in the Raman
  spectra of graphene-related materials}},\ }\href
  {https://doi.org/10.1088/2053-1583/aa5e77} {\bibfield  {journal} {\bibinfo
  {journal} {2D Materials}\ }\textbf {\bibinfo {volume} {4}},\ \bibinfo {pages}
  {025039} (\bibinfo {year} {2017})}\BibitemShut {NoStop}%
\bibitem [{\citenamefont {Gordeev}\ \emph {et~al.}(2016)\citenamefont
  {Gordeev}, \citenamefont {Setaro}, \citenamefont {Glaeske}, \citenamefont
  {J{\"{u}}rgensen},\ and\ \citenamefont {Reich}}]{Gordeev2016a}%
  \BibitemOpen
  \bibfield  {author} {\bibinfo {author} {\bibfnamefont {G.}~\bibnamefont
  {Gordeev}}, \bibinfo {author} {\bibfnamefont {A.}~\bibnamefont {Setaro}},
  \bibinfo {author} {\bibfnamefont {M.}~\bibnamefont {Glaeske}}, \bibinfo
  {author} {\bibfnamefont {S.}~\bibnamefont {J{\"{u}}rgensen}},\ and\ \bibinfo
  {author} {\bibfnamefont {S.}~\bibnamefont {Reich}},\ }\bibfield  {title}
  {\bibinfo {title} {{Doping in covalently functionalized carbon nanotubes: A
  Raman scattering study}},\ }\href {https://doi.org/10.1002/pssb.201600636}
  {\bibfield  {journal} {\bibinfo  {journal} {Physica Status Solidi (B)}\
  }\textbf {\bibinfo {volume} {253}},\ \bibinfo {pages} {2461} (\bibinfo {year}
  {2016})}\BibitemShut {NoStop}%
\bibitem [{\citenamefont {Grimm}\ \emph {et~al.}(2017)\citenamefont {Grimm},
  \citenamefont {Schie{\ss}l}, \citenamefont {Zakharko}, \citenamefont
  {Rother}, \citenamefont {Brohmann},\ and\ \citenamefont
  {Zaumseil}}]{Grimm2017}%
  \BibitemOpen
  \bibfield  {author} {\bibinfo {author} {\bibfnamefont {S.}~\bibnamefont
  {Grimm}}, \bibinfo {author} {\bibfnamefont {S.~P.}\ \bibnamefont
  {Schie{\ss}l}}, \bibinfo {author} {\bibfnamefont {Y.}~\bibnamefont
  {Zakharko}}, \bibinfo {author} {\bibfnamefont {M.}~\bibnamefont {Rother}},
  \bibinfo {author} {\bibfnamefont {M.}~\bibnamefont {Brohmann}},\ and\
  \bibinfo {author} {\bibfnamefont {J.}~\bibnamefont {Zaumseil}},\ }\bibfield
  {title} {\bibinfo {title} {{Doping-dependent G-mode shifts of small diameter
  semiconducting single-walled carbon nanotubes}},\ }\href
  {https://doi.org/10.1016/j.carbon.2017.03.040} {\bibfield  {journal}
  {\bibinfo  {journal} {Carbon}\ }\textbf {\bibinfo {volume} {118}},\ \bibinfo
  {pages} {261} (\bibinfo {year} {2017})}\BibitemShut {NoStop}%
\bibitem [{\citenamefont {Das}\ and\ \citenamefont {Sood}(2009)}]{Das2009}%
  \BibitemOpen
  \bibfield  {author} {\bibinfo {author} {\bibfnamefont {A.}~\bibnamefont
  {Das}}\ and\ \bibinfo {author} {\bibfnamefont {A.~K.}\ \bibnamefont {Sood}},\
  }\bibfield  {title} {\bibinfo {title} {{Renormalization of the phonon
  spectrum in semiconducting single-walled carbon nanotubes studied by Raman
  spectroscopy}},\ }\href {https://doi.org/10.1103/PhysRevB.79.235429}
  {\bibfield  {journal} {\bibinfo  {journal} {Physical Review B}\ }\textbf
  {\bibinfo {volume} {79}},\ \bibinfo {pages} {235429} (\bibinfo {year}
  {2009})}\BibitemShut {NoStop}%
\bibitem [{\citenamefont {Gaulke}\ \emph {et~al.}(2020)\citenamefont {Gaulke},
  \citenamefont {Janissek}, \citenamefont {Peyyety}, \citenamefont {Alamgir},
  \citenamefont {Riaz}, \citenamefont {Dehm}, \citenamefont {Li}, \citenamefont
  {Lemmer}, \citenamefont {Flavel}, \citenamefont {Kappes}, \citenamefont
  {Hennrich}, \citenamefont {Wei}, \citenamefont {Chen}, \citenamefont
  {Pyatkov},\ and\ \citenamefont {Krupke}}]{Gaulke2020}%
  \BibitemOpen
  \bibfield  {author} {\bibinfo {author} {\bibfnamefont {M.}~\bibnamefont
  {Gaulke}}, \bibinfo {author} {\bibfnamefont {A.}~\bibnamefont {Janissek}},
  \bibinfo {author} {\bibfnamefont {N.~A.}\ \bibnamefont {Peyyety}}, \bibinfo
  {author} {\bibfnamefont {I.}~\bibnamefont {Alamgir}}, \bibinfo {author}
  {\bibfnamefont {A.}~\bibnamefont {Riaz}}, \bibinfo {author} {\bibfnamefont
  {S.}~\bibnamefont {Dehm}}, \bibinfo {author} {\bibfnamefont {H.}~\bibnamefont
  {Li}}, \bibinfo {author} {\bibfnamefont {U.}~\bibnamefont {Lemmer}}, \bibinfo
  {author} {\bibfnamefont {B.~S.}\ \bibnamefont {Flavel}}, \bibinfo {author}
  {\bibfnamefont {M.~M.}\ \bibnamefont {Kappes}}, \bibinfo {author}
  {\bibfnamefont {F.}~\bibnamefont {Hennrich}}, \bibinfo {author}
  {\bibfnamefont {L.}~\bibnamefont {Wei}}, \bibinfo {author} {\bibfnamefont
  {Y.}~\bibnamefont {Chen}}, \bibinfo {author} {\bibfnamefont {F.}~\bibnamefont
  {Pyatkov}},\ and\ \bibinfo {author} {\bibfnamefont {R.}~\bibnamefont
  {Krupke}},\ }\bibfield  {title} {\bibinfo {title} {{Low temperature
  Electroluminescence Excitation Mapping of Excitons and Trions in
  Short-Channel Monochiral Carbon Nanotube Devices}},\ }\href
  {https://doi.org/10.1021/acsnano.9b07207} {\bibfield  {journal} {\bibinfo
  {journal} {ACS Nano}\ ,\ \bibinfo {pages} {2709–2717}} (\bibinfo {year}
  {2020})}\BibitemShut {NoStop}%
\bibitem [{\citenamefont {St{\"{u}}rzl}\ \emph {et~al.}(2009)\citenamefont
  {St{\"{u}}rzl}, \citenamefont {Hennrich}, \citenamefont {Lebedkin},\ and\
  \citenamefont {Kappes}}]{Sturzl2009}%
  \BibitemOpen
  \bibfield  {author} {\bibinfo {author} {\bibfnamefont {N.}~\bibnamefont
  {St{\"{u}}rzl}}, \bibinfo {author} {\bibfnamefont {F.}~\bibnamefont
  {Hennrich}}, \bibinfo {author} {\bibfnamefont {S.}~\bibnamefont {Lebedkin}},\
  and\ \bibinfo {author} {\bibfnamefont {M.~M.}\ \bibnamefont {Kappes}},\
  }\bibfield  {title} {\bibinfo {title} {{Near monochiral single-walled carbon
  nanotube dispersions in organic solvents}},\ }\href
  {https://doi.org/10.1021/jp902788y} {\bibfield  {journal} {\bibinfo
  {journal} {Journal of Physical Chemistry C}\ }\textbf {\bibinfo {volume}
  {113}},\ \bibinfo {pages} {14628} (\bibinfo {year} {2009})}\BibitemShut
  {NoStop}%
\bibitem [{\citenamefont {Hennrich}\ \emph {et~al.}(2016)\citenamefont
  {Hennrich}, \citenamefont {Li}, \citenamefont {Fischer}, \citenamefont
  {Lebedkin}, \citenamefont {Krupke},\ and\ \citenamefont
  {Kappes}}]{Hennrich2016}%
  \BibitemOpen
  \bibfield  {author} {\bibinfo {author} {\bibfnamefont {F.}~\bibnamefont
  {Hennrich}}, \bibinfo {author} {\bibfnamefont {W.}~\bibnamefont {Li}},
  \bibinfo {author} {\bibfnamefont {R.}~\bibnamefont {Fischer}}, \bibinfo
  {author} {\bibfnamefont {S.}~\bibnamefont {Lebedkin}}, \bibinfo {author}
  {\bibfnamefont {R.}~\bibnamefont {Krupke}},\ and\ \bibinfo {author}
  {\bibfnamefont {M.~M.}\ \bibnamefont {Kappes}},\ }\bibfield  {title}
  {\bibinfo {title} {{Length-Sorted, Large-Diameter, Polyfluorene-Wrapped
  Semiconducting Single-Walled Carbon Nanotubes for High-Density, Short-Channel
  Transistors}},\ }\href {https://doi.org/10.1021/acsnano.5b05572} {\bibfield
  {journal} {\bibinfo  {journal} {ACS Nano}\ }\textbf {\bibinfo {volume}
  {10}},\ \bibinfo {pages} {1888} (\bibinfo {year} {2016})}\BibitemShut
  {NoStop}%
\bibitem [{\citenamefont {Vijayaraghavan}\ \emph {et~al.}(2007)\citenamefont
  {Vijayaraghavan}, \citenamefont {Blatt}, \citenamefont {Weissenberger},
  \citenamefont {Oron-Carl}, \citenamefont {Hennrich}, \citenamefont
  {Gerthsen}, \citenamefont {Hahn},\ and\ \citenamefont
  {Krupke}}]{Vijayaraghavan2007}%
  \BibitemOpen
  \bibfield  {author} {\bibinfo {author} {\bibfnamefont {A.}~\bibnamefont
  {Vijayaraghavan}}, \bibinfo {author} {\bibfnamefont {S.}~\bibnamefont
  {Blatt}}, \bibinfo {author} {\bibfnamefont {D.}~\bibnamefont
  {Weissenberger}}, \bibinfo {author} {\bibfnamefont {M.}~\bibnamefont
  {Oron-Carl}}, \bibinfo {author} {\bibfnamefont {F.}~\bibnamefont {Hennrich}},
  \bibinfo {author} {\bibfnamefont {D.}~\bibnamefont {Gerthsen}}, \bibinfo
  {author} {\bibfnamefont {H.}~\bibnamefont {Hahn}},\ and\ \bibinfo {author}
  {\bibfnamefont {R.}~\bibnamefont {Krupke}},\ }\bibfield  {title} {\bibinfo
  {title} {{Ultra-large-scale directed assembly of single-walled carbon
  nanotube devices}},\ }\href {https://doi.org/10.1021/nl0703727} {\bibfield
  {journal} {\bibinfo  {journal} {Nano Letters}\ }\textbf {\bibinfo {volume}
  {7}},\ \bibinfo {pages} {1556} (\bibinfo {year} {2007})}\BibitemShut
  {NoStop}%
\bibitem [{\citenamefont {Tachibana}(2013)}]{Tachibana2013}%
  \BibitemOpen
  \bibfield  {author} {\bibinfo {author} {\bibfnamefont {M.}~\bibnamefont
  {Tachibana}},\ }\bibfield  {title} {\bibinfo {title} {{Characterization of
  Laser-Induced Defect Sand Modification in Carbon Nanotubes by Raman
  Spectroscopy}},\ }in\ \href {https://doi.org/10.5772/52091} {\emph {\bibinfo
  {booktitle} {Physical and Chemical Properties of Carbon Nanotubes}}},\
  \bibinfo {editor} {edited by\ \bibinfo {editor} {\bibfnamefont
  {S.}~\bibnamefont {Suzuki}}}\ (\bibinfo  {publisher} {InTech},\ \bibinfo
  {address} {Rijeka},\ \bibinfo {year} {2013})\ Chap.\ \bibinfo {chapter}
  {Characterization of Laser-Induced Defects and Modification in Carbon
  Nanotubes by Raman Spectroscopy}\BibitemShut {NoStop}%
\bibitem [{\citenamefont {Van}\ \emph {et~al.}(2017)\citenamefont {Van},
  \citenamefont {Badura}, \citenamefont {Liang}, \citenamefont {Okoli},\ and\
  \citenamefont {Zhang}}]{Van2017}%
  \BibitemOpen
  \bibfield  {author} {\bibinfo {author} {\bibfnamefont {H.~H.}\ \bibnamefont
  {Van}}, \bibinfo {author} {\bibfnamefont {K.}~\bibnamefont {Badura}},
  \bibinfo {author} {\bibfnamefont {R.}~\bibnamefont {Liang}}, \bibinfo
  {author} {\bibfnamefont {O.}~\bibnamefont {Okoli}},\ and\ \bibinfo {author}
  {\bibfnamefont {M.}~\bibnamefont {Zhang}},\ }\bibfield  {title} {\bibinfo
  {title} {{Laser-induced graphitic healing of carbon nanotubes aligned in a
  sheet}},\ }\href {https://doi.org/10.2351/1.4980166} {\bibfield  {journal}
  {\bibinfo  {journal} {Journal of Laser Applications}\ }\textbf {\bibinfo
  {volume} {29}},\ \bibinfo {pages} {022010} (\bibinfo {year}
  {2017})}\BibitemShut {NoStop}%
\bibitem [{\citenamefont {Setaro}\ \emph {et~al.}(2017)\citenamefont {Setaro},
  \citenamefont {Adeli}, \citenamefont {Glaeske}, \citenamefont {Przyrembel},
  \citenamefont {Bisswanger}, \citenamefont {Gordeev}, \citenamefont
  {Maschietto}, \citenamefont {Faghani}, \citenamefont {Paulus}, \citenamefont
  {Weinelt}, \citenamefont {Arenal}, \citenamefont {Haag},\ and\ \citenamefont
  {Reich}}]{Setaro2017}%
  \BibitemOpen
  \bibfield  {author} {\bibinfo {author} {\bibfnamefont {A.}~\bibnamefont
  {Setaro}}, \bibinfo {author} {\bibfnamefont {M.}~\bibnamefont {Adeli}},
  \bibinfo {author} {\bibfnamefont {M.}~\bibnamefont {Glaeske}}, \bibinfo
  {author} {\bibfnamefont {D.}~\bibnamefont {Przyrembel}}, \bibinfo {author}
  {\bibfnamefont {T.}~\bibnamefont {Bisswanger}}, \bibinfo {author}
  {\bibfnamefont {G.}~\bibnamefont {Gordeev}}, \bibinfo {author} {\bibfnamefont
  {F.}~\bibnamefont {Maschietto}}, \bibinfo {author} {\bibfnamefont
  {A.}~\bibnamefont {Faghani}}, \bibinfo {author} {\bibfnamefont
  {B.}~\bibnamefont {Paulus}}, \bibinfo {author} {\bibfnamefont
  {M.}~\bibnamefont {Weinelt}}, \bibinfo {author} {\bibfnamefont
  {R.}~\bibnamefont {Arenal}}, \bibinfo {author} {\bibfnamefont
  {R.}~\bibnamefont {Haag}},\ and\ \bibinfo {author} {\bibfnamefont
  {S.}~\bibnamefont {Reich}},\ }\bibfield  {title} {\bibinfo {title}
  {{Preserving $\pi$-conjugation in covalently functionalized carbon nanotubes
  for optoelectronic applications}},\ }\href
  {https://doi.org/10.1038/ncomms14281} {\bibfield  {journal} {\bibinfo
  {journal} {Nature Communications}\ }\textbf {\bibinfo {volume} {8}},\
  \bibinfo {pages} {14281} (\bibinfo {year} {2017})}\BibitemShut {NoStop}%
\bibitem [{\citenamefont {Connors}(1990)}]{Connors1990}%
  \BibitemOpen
  \bibfield  {author} {\bibinfo {author} {\bibfnamefont {A.~K.}\ \bibnamefont
  {Connors}},\ }\href@noop {} {\emph {\bibinfo {title} {Chemical kinetics: The
  study of reaction rates in solution}}},\ Vol.\ \bibinfo {volume} {180}\
  (\bibinfo  {publisher} {John Wiley {\&} Sons},\ \bibinfo {address} {New
  York},\ \bibinfo {year} {1990})\ p.\ \bibinfo {pages} {140}\BibitemShut
  {NoStop}%
\bibitem [{\citenamefont {Benfante}\ \emph {et~al.}(2018)\citenamefont
  {Benfante}, \citenamefont {Giambra}, \citenamefont {Pernice}, \citenamefont
  {Stivala}, \citenamefont {Calandra}, \citenamefont {Parisi}, \citenamefont
  {Cino}, \citenamefont {Dehm}, \citenamefont {Danneau}, \citenamefont
  {Krupke},\ and\ \citenamefont {Busacca}}]{Benfante2018}%
  \BibitemOpen
  \bibfield  {author} {\bibinfo {author} {\bibfnamefont {A.}~\bibnamefont
  {Benfante}}, \bibinfo {author} {\bibfnamefont {M.~A.}\ \bibnamefont
  {Giambra}}, \bibinfo {author} {\bibfnamefont {R.}~\bibnamefont {Pernice}},
  \bibinfo {author} {\bibfnamefont {S.}~\bibnamefont {Stivala}}, \bibinfo
  {author} {\bibfnamefont {E.}~\bibnamefont {Calandra}}, \bibinfo {author}
  {\bibfnamefont {A.}~\bibnamefont {Parisi}}, \bibinfo {author} {\bibfnamefont
  {A.~C.}\ \bibnamefont {Cino}}, \bibinfo {author} {\bibfnamefont
  {S.}~\bibnamefont {Dehm}}, \bibinfo {author} {\bibfnamefont {R.}~\bibnamefont
  {Danneau}}, \bibinfo {author} {\bibfnamefont {R.}~\bibnamefont {Krupke}},\
  and\ \bibinfo {author} {\bibfnamefont {A.~C.}\ \bibnamefont {Busacca}},\
  }\bibfield  {title} {\bibinfo {title} {{Employing Microwave Graphene Field
  Effect Transistors for Infrared Radiation Detection}},\ }\href
  {https://doi.org/10.1109/JPHOT.2018.2807923} {\bibfield  {journal} {\bibinfo
  {journal} {IEEE Photonics Journal}\ }\textbf {\bibinfo {volume} {10}},\
  \bibinfo {pages} {6801407} (\bibinfo {year} {2018})}\BibitemShut {NoStop}%
\bibitem [{\citenamefont {Riaz}\ \emph {et~al.}(2019)\citenamefont {Riaz},
  \citenamefont {Alam}, \citenamefont {Selvasundaram}, \citenamefont {Dehm},
  \citenamefont {Hennrich}, \citenamefont {Kappes},\ and\ \citenamefont
  {Krupke}}]{Riaz2019}%
  \BibitemOpen
  \bibfield  {author} {\bibinfo {author} {\bibfnamefont {A.}~\bibnamefont
  {Riaz}}, \bibinfo {author} {\bibfnamefont {A.}~\bibnamefont {Alam}}, \bibinfo
  {author} {\bibfnamefont {P.~B.}\ \bibnamefont {Selvasundaram}}, \bibinfo
  {author} {\bibfnamefont {S.}~\bibnamefont {Dehm}}, \bibinfo {author}
  {\bibfnamefont {F.}~\bibnamefont {Hennrich}}, \bibinfo {author}
  {\bibfnamefont {M.~M.}\ \bibnamefont {Kappes}},\ and\ \bibinfo {author}
  {\bibfnamefont {R.}~\bibnamefont {Krupke}},\ }\bibfield  {title} {\bibinfo
  {title} {{Near-Infrared Photoresponse of Waveguide-Integrated Carbon
  Nanotube–Silicon Junctions}},\ }\href
  {https://doi.org/10.1002/aelm.201800265} {\bibfield  {journal} {\bibinfo
  {journal} {Advanced Electronic Materials}\ }\textbf {\bibinfo {volume} {5}},\
  \bibinfo {pages} {1800265} (\bibinfo {year} {2019})}\BibitemShut {NoStop}%
\bibitem [{\citenamefont {Schmidt}\ \emph {et~al.}(2009)\citenamefont
  {Schmidt}, \citenamefont {Gallon}, \citenamefont {Esnouf}, \citenamefont
  {Bourgoin},\ and\ \citenamefont {Chenevier}}]{Schmidt2009}%
  \BibitemOpen
  \bibfield  {author} {\bibinfo {author} {\bibfnamefont {G.}~\bibnamefont
  {Schmidt}}, \bibinfo {author} {\bibfnamefont {S.}~\bibnamefont {Gallon}},
  \bibinfo {author} {\bibfnamefont {S.}~\bibnamefont {Esnouf}}, \bibinfo
  {author} {\bibfnamefont {J.~P.}\ \bibnamefont {Bourgoin}},\ and\ \bibinfo
  {author} {\bibfnamefont {P.}~\bibnamefont {Chenevier}},\ }\bibfield  {title}
  {\bibinfo {title} {{Mechanism of the coupling of diazonium to single-walled
  carbon nanotubes and its consequences}},\ }\href
  {https://doi.org/10.1002/chem.200801801} {\bibfield  {journal} {\bibinfo
  {journal} {Chemistry - A European Journal}\ }\textbf {\bibinfo {volume}
  {15}},\ \bibinfo {pages} {2101} (\bibinfo {year} {2009})}\BibitemShut
  {NoStop}%
\bibitem [{\citenamefont {Dyke}\ \emph {et~al.}(2004)\citenamefont {Dyke},
  \citenamefont {Stewart}, \citenamefont {Maya},\ and\ \citenamefont
  {Tour}}]{Dyke2004}%
  \BibitemOpen
  \bibfield  {author} {\bibinfo {author} {\bibfnamefont {C.~A.}\ \bibnamefont
  {Dyke}}, \bibinfo {author} {\bibfnamefont {M.~P.}\ \bibnamefont {Stewart}},
  \bibinfo {author} {\bibfnamefont {F.}~\bibnamefont {Maya}},\ and\ \bibinfo
  {author} {\bibfnamefont {J.~M.}\ \bibnamefont {Tour}},\ }\bibfield  {title}
  {\bibinfo {title} {{Diazonium-Based Functionalization of Carbon Nanotubes:
  XPS and GC-MS Analysis and Mechanistic Implications}},\ }\href
  {https://doi.org/10.1055/s-2003-44983} {\bibfield  {journal} {\bibinfo
  {journal} {Synlett}\ ,\ \bibinfo {pages} {155}} (\bibinfo {year}
  {2004})}\BibitemShut {NoStop}%
\bibitem [{\citenamefont {Galli}(1988)}]{Galli1988}%
  \BibitemOpen
  \bibfield  {author} {\bibinfo {author} {\bibfnamefont {C.}~\bibnamefont
  {Galli}},\ }\bibfield  {title} {\bibinfo {title} {{Radical Reactions of
  Arenediazonium Ions: An Easy Entry into the Chemistry of the Aryl Radical}},\
  }\href {https://doi.org/10.1021/cr00087a004} {\bibfield  {journal} {\bibinfo
  {journal} {Chemical Reviews}\ }\textbf {\bibinfo {volume} {88}},\ \bibinfo
  {pages} {765} (\bibinfo {year} {1988})}\BibitemShut {NoStop}%
\bibitem [{\citenamefont {Gomberg}\ and\ \citenamefont
  {Bachmann}(1924)}]{Gomberg1924}%
  \BibitemOpen
  \bibfield  {author} {\bibinfo {author} {\bibfnamefont {M.}~\bibnamefont
  {Gomberg}}\ and\ \bibinfo {author} {\bibfnamefont {W.~E.}\ \bibnamefont
  {Bachmann}},\ }\bibfield  {title} {\bibinfo {title} {{The synthesis of biaryl
  compounds by means of the diazo reaction}},\ }\href
  {https://doi.org/10.1021/ja01675a026} {\bibfield  {journal} {\bibinfo
  {journal} {Journal of the American Chemical Society}\ }\textbf {\bibinfo
  {volume} {46}},\ \bibinfo {pages} {2339} (\bibinfo {year}
  {1924})}\BibitemShut {NoStop}%
\bibitem [{\citenamefont {Burek}\ \emph {et~al.}(2019)\citenamefont {Burek},
  \citenamefont {Bahnemann},\ and\ \citenamefont {Bloh}}]{Burek2019}%
  \BibitemOpen
  \bibfield  {author} {\bibinfo {author} {\bibfnamefont {B.~O.}\ \bibnamefont
  {Burek}}, \bibinfo {author} {\bibfnamefont {D.~W.}\ \bibnamefont
  {Bahnemann}},\ and\ \bibinfo {author} {\bibfnamefont {J.~Z.}\ \bibnamefont
  {Bloh}},\ }\bibfield  {title} {\bibinfo {title} {{Modeling and Optimization
  of the Photocatalytic Reduction of Molecular Oxygen to Hydrogen Peroxide over
  Titanium Dioxide}},\ }\href {https://doi.org/10.1021/acscatal.8b03638}
  {\bibfield  {journal} {\bibinfo  {journal} {ACS Catalysis}\ }\textbf
  {\bibinfo {volume} {9}},\ \bibinfo {pages} {25} (\bibinfo {year}
  {2019})}\BibitemShut {NoStop}%
\bibitem [{\citenamefont {DeTar}\ and\ \citenamefont {{Ray
  Ballentine}}(1956)}]{DeTar1956}%
  \BibitemOpen
  \bibfield  {author} {\bibinfo {author} {\bibfnamefont {D.~L.~F.}\
  \bibnamefont {DeTar}}\ and\ \bibinfo {author} {\bibfnamefont
  {A.}~\bibnamefont {{Ray Ballentine}}},\ }\bibfield  {title} {\bibinfo {title}
  {{The Mechanisms of Diazonium Salt Reactions. II. A Redetermination of the
  Rates of the Thermal Decomposition of Six Diazonium Salts in Aqueous
  Solution}},\ }\href {https://doi.org/10.1021/ja01597a015} {\bibfield
  {journal} {\bibinfo  {journal} {Journal of the American Chemical Society}\
  }\textbf {\bibinfo {volume} {78}},\ \bibinfo {pages} {3916} (\bibinfo {year}
  {1956})}\BibitemShut {NoStop}%
\bibitem [{\citenamefont {Hatting}\ \emph {et~al.}(2013)\citenamefont
  {Hatting}, \citenamefont {Heeg}, \citenamefont {Ataka}, \citenamefont
  {Heberle}, \citenamefont {Hennrich}, \citenamefont {Kappes}, \citenamefont
  {Krupke},\ and\ \citenamefont {Reich}}]{Hatting2013}%
  \BibitemOpen
  \bibfield  {author} {\bibinfo {author} {\bibfnamefont {B.}~\bibnamefont
  {Hatting}}, \bibinfo {author} {\bibfnamefont {S.}~\bibnamefont {Heeg}},
  \bibinfo {author} {\bibfnamefont {K.}~\bibnamefont {Ataka}}, \bibinfo
  {author} {\bibfnamefont {J.}~\bibnamefont {Heberle}}, \bibinfo {author}
  {\bibfnamefont {F.}~\bibnamefont {Hennrich}}, \bibinfo {author}
  {\bibfnamefont {M.~M.}\ \bibnamefont {Kappes}}, \bibinfo {author}
  {\bibfnamefont {R.}~\bibnamefont {Krupke}},\ and\ \bibinfo {author}
  {\bibfnamefont {S.}~\bibnamefont {Reich}},\ }\bibfield  {title} {\bibinfo
  {title} {{Fermi energy shift in deposited metallic nanotubes: A Raman
  scattering study}},\ }\href {https://doi.org/10.1103/PhysRevB.87.165442}
  {\bibfield  {journal} {\bibinfo  {journal} {Physical Review B}\ }\textbf
  {\bibinfo {volume} {87}},\ \bibinfo {pages} {165442} (\bibinfo {year}
  {2013})}\BibitemShut {NoStop}%
\bibitem [{\citenamefont {Maultzsch}\ \emph {et~al.}(2005)\citenamefont
  {Maultzsch}, \citenamefont {Telg}, \citenamefont {Reich},\ and\ \citenamefont
  {Thomsen}}]{Maultzsch2005a}%
  \BibitemOpen
  \bibfield  {author} {\bibinfo {author} {\bibfnamefont {J.}~\bibnamefont
  {Maultzsch}}, \bibinfo {author} {\bibfnamefont {H.}~\bibnamefont {Telg}},
  \bibinfo {author} {\bibfnamefont {S.}~\bibnamefont {Reich}},\ and\ \bibinfo
  {author} {\bibfnamefont {C.}~\bibnamefont {Thomsen}},\ }\bibfield  {title}
  {\bibinfo {title} {{Radial breathing mode of single-walled carbon nanotubes:
  Optical transition energies and chiral-index assignment}},\ }\href
  {https://doi.org/10.1103/PhysRevB.72.205438} {\bibfield  {journal} {\bibinfo
  {journal} {Physical Review B}\ }\textbf {\bibinfo {volume} {72}},\ \bibinfo
  {pages} {205438} (\bibinfo {year} {2005})},\ \Eprint
  {https://arxiv.org/abs/0510427} {arXiv:0510427 [cond-mat]} \BibitemShut
  {NoStop}%
\bibitem [{\citenamefont {Gordeev}\ \emph {et~al.}(2017)\citenamefont
  {Gordeev}, \citenamefont {Jorio}, \citenamefont {Kusch}, \citenamefont
  {Vieira}, \citenamefont {Flavel}, \citenamefont {Krupke}, \citenamefont
  {Barros},\ and\ \citenamefont {Reich}}]{Gordeev2017}%
  \BibitemOpen
  \bibfield  {author} {\bibinfo {author} {\bibfnamefont {G.}~\bibnamefont
  {Gordeev}}, \bibinfo {author} {\bibfnamefont {A.}~\bibnamefont {Jorio}},
  \bibinfo {author} {\bibfnamefont {P.}~\bibnamefont {Kusch}}, \bibinfo
  {author} {\bibfnamefont {B.~G.}\ \bibnamefont {Vieira}}, \bibinfo {author}
  {\bibfnamefont {B.}~\bibnamefont {Flavel}}, \bibinfo {author} {\bibfnamefont
  {R.}~\bibnamefont {Krupke}}, \bibinfo {author} {\bibfnamefont {E.~B.}\
  \bibnamefont {Barros}},\ and\ \bibinfo {author} {\bibfnamefont
  {S.}~\bibnamefont {Reich}},\ }\bibfield  {title} {\bibinfo {title} {{Resonant
  anti-Stokes Raman scattering in single-walled carbon nanotubes}},\ }\href
  {https://doi.org/10.1103/PhysRevB.96.245415} {\bibfield  {journal} {\bibinfo
  {journal} {Physical Review B}\ }\textbf {\bibinfo {volume} {96}},\ \bibinfo
  {pages} {245415} (\bibinfo {year} {2017})}\BibitemShut {NoStop}%
\bibitem [{\citenamefont {Perebeinos}\ \emph {et~al.}(2004)\citenamefont
  {Perebeinos}, \citenamefont {Tersoff},\ and\ \citenamefont
  {Avouris}}]{Perebeinos2004}%
  \BibitemOpen
  \bibfield  {author} {\bibinfo {author} {\bibfnamefont {V.}~\bibnamefont
  {Perebeinos}}, \bibinfo {author} {\bibfnamefont {J.}~\bibnamefont
  {Tersoff}},\ and\ \bibinfo {author} {\bibfnamefont {P.}~\bibnamefont
  {Avouris}},\ }\bibfield  {title} {\bibinfo {title} {{Scaling of excitons in
  carbon nanotubes}},\ }\href {https://doi.org/10.1103/PhysRevLett.92.257402}
  {\bibfield  {journal} {\bibinfo  {journal} {Physical Review Letters}\
  }\textbf {\bibinfo {volume} {92}},\ \bibinfo {pages} {257402} (\bibinfo
  {year} {2004})},\ \Eprint {https://arxiv.org/abs/0402091} {arXiv:0402091
  [cond-mat]} \BibitemShut {NoStop}%
\bibitem [{\citenamefont {Walsh}\ \emph {et~al.}(2007)\citenamefont {Walsh},
  \citenamefont {Vamivakas}, \citenamefont {Yin}, \citenamefont {Cronin},
  \citenamefont {{\"{U}}nl{\"{u}}}, \citenamefont {Goldberg},\ and\
  \citenamefont {Swan}}]{Walsh2007}%
  \BibitemOpen
  \bibfield  {author} {\bibinfo {author} {\bibfnamefont {A.~G.}\ \bibnamefont
  {Walsh}}, \bibinfo {author} {\bibfnamefont {A.~N.}\ \bibnamefont
  {Vamivakas}}, \bibinfo {author} {\bibfnamefont {Y.}~\bibnamefont {Yin}},
  \bibinfo {author} {\bibfnamefont {S.~B.}\ \bibnamefont {Cronin}}, \bibinfo
  {author} {\bibfnamefont {M.~S.}\ \bibnamefont {{\"{U}}nl{\"{u}}}}, \bibinfo
  {author} {\bibfnamefont {B.~B.}\ \bibnamefont {Goldberg}},\ and\ \bibinfo
  {author} {\bibfnamefont {A.~K.}\ \bibnamefont {Swan}},\ }\bibfield  {title}
  {\bibinfo {title} {{Screening of excitons in single, suspended carbon
  nanotubes}},\ }\href {https://doi.org/10.1021/nl070193p} {\bibfield
  {journal} {\bibinfo  {journal} {Nano Letters}\ }\textbf {\bibinfo {volume}
  {7}},\ \bibinfo {pages} {1485} (\bibinfo {year} {2007})}\BibitemShut
  {NoStop}%
\bibitem [{\citenamefont {Araujo}\ \emph {et~al.}(2009)\citenamefont {Araujo},
  \citenamefont {Jorio}, \citenamefont {Dresselhaus}, \citenamefont {Sato},\
  and\ \citenamefont {Saito}}]{Araujo2009}%
  \BibitemOpen
  \bibfield  {author} {\bibinfo {author} {\bibfnamefont {P.~T.}\ \bibnamefont
  {Araujo}}, \bibinfo {author} {\bibfnamefont {A.}~\bibnamefont {Jorio}},
  \bibinfo {author} {\bibfnamefont {M.~S.}\ \bibnamefont {Dresselhaus}},
  \bibinfo {author} {\bibfnamefont {K.}~\bibnamefont {Sato}},\ and\ \bibinfo
  {author} {\bibfnamefont {R.}~\bibnamefont {Saito}},\ }\bibfield  {title}
  {\bibinfo {title} {{Diameter dependence of the dielectric constant for the
  excitonic transition energy of single-wall carbon nanotubes}},\ }\href
  {https://doi.org/10.1103/PhysRevLett.103.146802} {\bibfield  {journal}
  {\bibinfo  {journal} {Physical Review Letters}\ }\textbf {\bibinfo {volume}
  {103}},\ \bibinfo {pages} {146802} (\bibinfo {year} {2009})}\BibitemShut
  {NoStop}%
\bibitem [{\citenamefont {Roquelet}\ \emph {et~al.}(2010)\citenamefont
  {Roquelet}, \citenamefont {Lauret}, \citenamefont {Alain-Rizzo},
  \citenamefont {Voisin}, \citenamefont {Fleurier}, \citenamefont {Delarue},
  \citenamefont {Garrot}, \citenamefont {Loiseau}, \citenamefont {Roussignol},
  \citenamefont {Delaire},\ and\ \citenamefont {Deleporte}}]{Roquelet2010}%
  \BibitemOpen
  \bibfield  {author} {\bibinfo {author} {\bibfnamefont {C.}~\bibnamefont
  {Roquelet}}, \bibinfo {author} {\bibfnamefont {J.~S.}\ \bibnamefont
  {Lauret}}, \bibinfo {author} {\bibfnamefont {V.}~\bibnamefont {Alain-Rizzo}},
  \bibinfo {author} {\bibfnamefont {C.}~\bibnamefont {Voisin}}, \bibinfo
  {author} {\bibfnamefont {R.}~\bibnamefont {Fleurier}}, \bibinfo {author}
  {\bibfnamefont {M.}~\bibnamefont {Delarue}}, \bibinfo {author} {\bibfnamefont
  {D.}~\bibnamefont {Garrot}}, \bibinfo {author} {\bibfnamefont
  {A.}~\bibnamefont {Loiseau}}, \bibinfo {author} {\bibfnamefont
  {P.}~\bibnamefont {Roussignol}}, \bibinfo {author} {\bibfnamefont {J.~A.}\
  \bibnamefont {Delaire}},\ and\ \bibinfo {author} {\bibfnamefont
  {E.}~\bibnamefont {Deleporte}},\ }\bibfield  {title} {\bibinfo {title}
  {{$\Pi$-Stacking functionalization of carbon nanotubes through micelle
  swelling}},\ }\href {https://doi.org/10.1002/cphc.201000067} {\bibfield
  {journal} {\bibinfo  {journal} {ChemPhysChem}\ }\textbf {\bibinfo {volume}
  {11}},\ \bibinfo {pages} {1667} (\bibinfo {year} {2010})}\BibitemShut
  {NoStop}%
\bibitem [{\citenamefont {Marquardt}\ \emph {et~al.}(2010)\citenamefont
  {Marquardt}, \citenamefont {Grunder}, \citenamefont {B{\l}aszczyk},
  \citenamefont {Dehm}, \citenamefont {Hennrich}, \citenamefont
  {L{\"{o}}hneysen}, \citenamefont {Mayor},\ and\ \citenamefont
  {Krupke}}]{Marquardt2010}%
  \BibitemOpen
  \bibfield  {author} {\bibinfo {author} {\bibfnamefont {C.~W.}\ \bibnamefont
  {Marquardt}}, \bibinfo {author} {\bibfnamefont {S.}~\bibnamefont {Grunder}},
  \bibinfo {author} {\bibfnamefont {A.}~\bibnamefont {B{\l}aszczyk}}, \bibinfo
  {author} {\bibfnamefont {S.}~\bibnamefont {Dehm}}, \bibinfo {author}
  {\bibfnamefont {F.}~\bibnamefont {Hennrich}}, \bibinfo {author}
  {\bibfnamefont {H.~V.}\ \bibnamefont {L{\"{o}}hneysen}}, \bibinfo {author}
  {\bibfnamefont {M.}~\bibnamefont {Mayor}},\ and\ \bibinfo {author}
  {\bibfnamefont {R.}~\bibnamefont {Krupke}},\ }\bibfield  {title} {\bibinfo
  {title} {{Electroluminescence from a single nanotube-molecule-nanotube
  junction}},\ }\href {https://doi.org/10.1038/nnano.2010.230} {\bibfield
  {journal} {\bibinfo  {journal} {Nature Nanotechnology}\ }\textbf {\bibinfo
  {volume} {5}},\ \bibinfo {pages} {863} (\bibinfo {year} {2010})}\BibitemShut
  {NoStop}%
\bibitem [{\citenamefont {Bachilo}\ \emph {et~al.}(2002)\citenamefont
  {Bachilo}, \citenamefont {Strano}, \citenamefont {Kittrell}, \citenamefont
  {Hauge}, \citenamefont {Smalley},\ and\ \citenamefont
  {Weisman}}]{Bachilo2002}%
  \BibitemOpen
  \bibfield  {author} {\bibinfo {author} {\bibfnamefont {S.~M.}\ \bibnamefont
  {Bachilo}}, \bibinfo {author} {\bibfnamefont {M.~S.}\ \bibnamefont {Strano}},
  \bibinfo {author} {\bibfnamefont {C.}~\bibnamefont {Kittrell}}, \bibinfo
  {author} {\bibfnamefont {R.~H.}\ \bibnamefont {Hauge}}, \bibinfo {author}
  {\bibfnamefont {R.~E.}\ \bibnamefont {Smalley}},\ and\ \bibinfo {author}
  {\bibfnamefont {R.~B.}\ \bibnamefont {Weisman}},\ }\bibfield  {title}
  {\bibinfo {title} {{Structure-assigned optical spectra of single-walled
  carbon nanotubes}},\ }\href {https://doi.org/10.1126/science.1078727}
  {\bibfield  {journal} {\bibinfo  {journal} {Science}\ }\textbf {\bibinfo
  {volume} {298}},\ \bibinfo {pages} {2361} (\bibinfo {year}
  {2002})}\BibitemShut {NoStop}%
\bibitem [{\citenamefont {Greben}\ \emph {et~al.}(2020)\citenamefont {Greben},
  \citenamefont {Arora}, \citenamefont {Harats},\ and\ \citenamefont
  {Bolotin}}]{Greben2020}%
  \BibitemOpen
  \bibfield  {author} {\bibinfo {author} {\bibfnamefont {K.}~\bibnamefont
  {Greben}}, \bibinfo {author} {\bibfnamefont {S.}~\bibnamefont {Arora}},
  \bibinfo {author} {\bibfnamefont {M.~G.}\ \bibnamefont {Harats}},\ and\
  \bibinfo {author} {\bibfnamefont {K.~I.}\ \bibnamefont {Bolotin}},\
  }\bibfield  {title} {\bibinfo {title} {{Intrinsic and Extrinsic
  Defect-Related Excitons in TMDCs}},\ }\href
  {https://doi.org/10.1021/acs.nanolett.9b05323} {\bibfield  {journal}
  {\bibinfo  {journal} {Nano Letters}\ }\textbf {\bibinfo {volume} {20}},\
  \bibinfo {pages} {2544} (\bibinfo {year} {2020})},\ \Eprint
  {https://arxiv.org/abs/1912.12327} {arXiv:1912.12327} \BibitemShut {NoStop}%
\bibitem [{\citenamefont {Wilson}\ \emph {et~al.}(2016)\citenamefont {Wilson},
  \citenamefont {Ripp}, \citenamefont {Prisbrey}, \citenamefont {Brown},
  \citenamefont {Sharf}, \citenamefont {Myles}, \citenamefont {Blank},\ and\
  \citenamefont {Minot}}]{Wilson2016}%
  \BibitemOpen
  \bibfield  {author} {\bibinfo {author} {\bibfnamefont {H.}~\bibnamefont
  {Wilson}}, \bibinfo {author} {\bibfnamefont {S.}~\bibnamefont {Ripp}},
  \bibinfo {author} {\bibfnamefont {L.}~\bibnamefont {Prisbrey}}, \bibinfo
  {author} {\bibfnamefont {M.~A.}\ \bibnamefont {Brown}}, \bibinfo {author}
  {\bibfnamefont {T.}~\bibnamefont {Sharf}}, \bibinfo {author} {\bibfnamefont
  {D.~J.}\ \bibnamefont {Myles}}, \bibinfo {author} {\bibfnamefont {K.~G.}\
  \bibnamefont {Blank}},\ and\ \bibinfo {author} {\bibfnamefont {E.~D.}\
  \bibnamefont {Minot}},\ }\bibfield  {title} {\bibinfo {title} {{Electrical
  monitoring of sp3 defect formation in individual carbon nanotubes}},\ }\href
  {https://doi.org/10.1021/acs.jpcc.5b11272} {\bibfield  {journal} {\bibinfo
  {journal} {Journal of Physical Chemistry C}\ }\textbf {\bibinfo {volume}
  {120}},\ \bibinfo {pages} {1971} (\bibinfo {year} {2016})}\BibitemShut
  {NoStop}%
\bibitem [{\citenamefont {Lee}\ \emph {et~al.}(2018)\citenamefont {Lee},
  \citenamefont {Trocchia}, \citenamefont {Warren}, \citenamefont {Young},
  \citenamefont {Vernick},\ and\ \citenamefont {Shepard}}]{Lee2018}%
  \BibitemOpen
  \bibfield  {author} {\bibinfo {author} {\bibfnamefont {Y.}~\bibnamefont
  {Lee}}, \bibinfo {author} {\bibfnamefont {S.~M.}\ \bibnamefont {Trocchia}},
  \bibinfo {author} {\bibfnamefont {S.~B.}\ \bibnamefont {Warren}}, \bibinfo
  {author} {\bibfnamefont {E.~F.}\ \bibnamefont {Young}}, \bibinfo {author}
  {\bibfnamefont {S.}~\bibnamefont {Vernick}},\ and\ \bibinfo {author}
  {\bibfnamefont {K.~L.}\ \bibnamefont {Shepard}},\ }\bibfield  {title}
  {\bibinfo {title} {{Electrically Controllable Single-Point Covalent
  Functionalization of Spin-Cast Carbon-Nanotube Field-Effect Transistor
  Arrays}},\ }\href {https://doi.org/10.1021/acsnano.8b03073} {\bibfield
  {journal} {\bibinfo  {journal} {ACS Nano}\ }\textbf {\bibinfo {volume}
  {12}},\ \bibinfo {pages} {9922–9930} (\bibinfo {year} {2018})}\BibitemShut
  {NoStop}%
\bibitem [{\citenamefont {Goldsmith}\ \emph {et~al.}(2007)\citenamefont
  {Goldsmith}, \citenamefont {Coroneus}, \citenamefont {Khalap}, \citenamefont
  {Kane}, \citenamefont {Weiss},\ and\ \citenamefont
  {Collins}}]{Goldsmith2007}%
  \BibitemOpen
  \bibfield  {author} {\bibinfo {author} {\bibfnamefont {B.~R.}\ \bibnamefont
  {Goldsmith}}, \bibinfo {author} {\bibfnamefont {J.~G.}\ \bibnamefont
  {Coroneus}}, \bibinfo {author} {\bibfnamefont {V.~R.}\ \bibnamefont
  {Khalap}}, \bibinfo {author} {\bibfnamefont {A.~A.}\ \bibnamefont {Kane}},
  \bibinfo {author} {\bibfnamefont {G.~A.}\ \bibnamefont {Weiss}},\ and\
  \bibinfo {author} {\bibfnamefont {P.~G.}\ \bibnamefont {Collins}},\
  }\bibfield  {title} {\bibinfo {title} {{Conductance-controlled point
  functionalization of single-walled carbon nanotubes}},\ }\href
  {https://doi.org/10.1126/science.1135303} {\bibfield  {journal} {\bibinfo
  {journal} {Science}\ }\textbf {\bibinfo {volume} {315}},\ \bibinfo {pages}
  {77} (\bibinfo {year} {2007})}\BibitemShut {NoStop}%
\bibitem [{\citenamefont {Berger}\ \emph {et~al.}(2009)\citenamefont {Berger},
  \citenamefont {Iglesias}, \citenamefont {Bonnet}, \citenamefont {Voisin},
  \citenamefont {Cassabois}, \citenamefont {Lauret}, \citenamefont
  {Delalande},\ and\ \citenamefont {Roussignol}}]{Berger2009}%
  \BibitemOpen
  \bibfield  {author} {\bibinfo {author} {\bibfnamefont {S.}~\bibnamefont
  {Berger}}, \bibinfo {author} {\bibfnamefont {F.}~\bibnamefont {Iglesias}},
  \bibinfo {author} {\bibfnamefont {P.}~\bibnamefont {Bonnet}}, \bibinfo
  {author} {\bibfnamefont {C.}~\bibnamefont {Voisin}}, \bibinfo {author}
  {\bibfnamefont {G.}~\bibnamefont {Cassabois}}, \bibinfo {author}
  {\bibfnamefont {J.~S.}\ \bibnamefont {Lauret}}, \bibinfo {author}
  {\bibfnamefont {C.}~\bibnamefont {Delalande}},\ and\ \bibinfo {author}
  {\bibfnamefont {P.}~\bibnamefont {Roussignol}},\ }\bibfield  {title}
  {\bibinfo {title} {{Optical properties of carbon nanotubes in a composite
  material: The role of dielectric screening and thermal expansion}},\ }\href
  {https://doi.org/10.1063/1.3116723} {\bibfield  {journal} {\bibinfo
  {journal} {Journal of Applied Physics}\ }\textbf {\bibinfo {volume} {105}},\
  \bibinfo {pages} {94323} (\bibinfo {year} {2009})}\BibitemShut {NoStop}%
\bibitem [{\citenamefont {Li}\ \emph {et~al.}(2019)\citenamefont {Li},
  \citenamefont {Gordeev}, \citenamefont {Garrity}, \citenamefont {Reich},\
  and\ \citenamefont {Flavel}}]{Li2019}%
  \BibitemOpen
  \bibfield  {author} {\bibinfo {author} {\bibfnamefont {H.}~\bibnamefont
  {Li}}, \bibinfo {author} {\bibfnamefont {G.}~\bibnamefont {Gordeev}},
  \bibinfo {author} {\bibfnamefont {O.}~\bibnamefont {Garrity}}, \bibinfo
  {author} {\bibfnamefont {S.}~\bibnamefont {Reich}},\ and\ \bibinfo {author}
  {\bibfnamefont {B.~S.}\ \bibnamefont {Flavel}},\ }\bibfield  {title}
  {\bibinfo {title} {{Separation of Small-Diameter Single-Walled Carbon
  Nanotubes in One to Three Steps with Aqueous Two-Phase Extraction}},\ }\href
  {https://doi.org/10.1021/acsnano.8b09579} {\bibfield  {journal} {\bibinfo
  {journal} {ACS Nano}\ }\textbf {\bibinfo {volume} {13}},\ \bibinfo {pages}
  {2567} (\bibinfo {year} {2019})}\BibitemShut {NoStop}%
\bibitem [{\citenamefont {Graf}\ \emph {et~al.}(2016)\citenamefont {Graf},
  \citenamefont {Zakharko}, \citenamefont {Schie{\ss}l}, \citenamefont
  {Backes}, \citenamefont {Pfohl}, \citenamefont {Flavel},\ and\ \citenamefont
  {Zaumseil}}]{Graf2016}%
  \BibitemOpen
  \bibfield  {author} {\bibinfo {author} {\bibfnamefont {A.}~\bibnamefont
  {Graf}}, \bibinfo {author} {\bibfnamefont {Y.}~\bibnamefont {Zakharko}},
  \bibinfo {author} {\bibfnamefont {S.~P.}\ \bibnamefont {Schie{\ss}l}},
  \bibinfo {author} {\bibfnamefont {C.}~\bibnamefont {Backes}}, \bibinfo
  {author} {\bibfnamefont {M.}~\bibnamefont {Pfohl}}, \bibinfo {author}
  {\bibfnamefont {B.~S.}\ \bibnamefont {Flavel}},\ and\ \bibinfo {author}
  {\bibfnamefont {J.}~\bibnamefont {Zaumseil}},\ }\bibfield  {title} {\bibinfo
  {title} {{Large scale, selective dispersion of long single-walled carbon
  nanotubes with high photoluminescence quantum yield by shear force mixing}},\
  }\href {https://doi.org/10.1016/j.carbon.2016.05.002} {\bibfield  {journal}
  {\bibinfo  {journal} {Carbon}\ }\textbf {\bibinfo {volume} {105}},\ \bibinfo
  {pages} {593} (\bibinfo {year} {2016})}\BibitemShut {NoStop}%
\bibitem [{\citenamefont {Denk}\ \emph {et~al.}(2014)\citenamefont {Denk},
  \citenamefont {Hohage}, \citenamefont {Zeppenfeld}, \citenamefont {Cai},
  \citenamefont {Pignedoli}, \citenamefont {S{\"{o}}de}, \citenamefont {Fasel},
  \citenamefont {Feng}, \citenamefont {M{\"{u}}llen}, \citenamefont {Wang},
  \citenamefont {Prezzi}, \citenamefont {Ferretti}, \citenamefont {Ruini},
  \citenamefont {Molinari},\ and\ \citenamefont {Ruffieux}}]{Denk2014}%
  \BibitemOpen
  \bibfield  {author} {\bibinfo {author} {\bibfnamefont {R.}~\bibnamefont
  {Denk}}, \bibinfo {author} {\bibfnamefont {M.}~\bibnamefont {Hohage}},
  \bibinfo {author} {\bibfnamefont {P.}~\bibnamefont {Zeppenfeld}}, \bibinfo
  {author} {\bibfnamefont {J.}~\bibnamefont {Cai}}, \bibinfo {author}
  {\bibfnamefont {C.~A.}\ \bibnamefont {Pignedoli}}, \bibinfo {author}
  {\bibfnamefont {H.}~\bibnamefont {S{\"{o}}de}}, \bibinfo {author}
  {\bibfnamefont {R.}~\bibnamefont {Fasel}}, \bibinfo {author} {\bibfnamefont
  {X.}~\bibnamefont {Feng}}, \bibinfo {author} {\bibfnamefont {K.}~\bibnamefont
  {M{\"{u}}llen}}, \bibinfo {author} {\bibfnamefont {S.}~\bibnamefont {Wang}},
  \bibinfo {author} {\bibfnamefont {D.}~\bibnamefont {Prezzi}}, \bibinfo
  {author} {\bibfnamefont {A.}~\bibnamefont {Ferretti}}, \bibinfo {author}
  {\bibfnamefont {A.}~\bibnamefont {Ruini}}, \bibinfo {author} {\bibfnamefont
  {E.}~\bibnamefont {Molinari}},\ and\ \bibinfo {author} {\bibfnamefont
  {P.}~\bibnamefont {Ruffieux}},\ }\bibfield  {title} {\bibinfo {title}
  {{Exciton-dominated optical response of ultra-narrow graphene nanoribbons}},\
  }\href {https://doi.org/10.1038/ncomms5253} {\bibfield  {journal} {\bibinfo
  {journal} {Nature Communications}\ }\textbf {\bibinfo {volume} {5}},\
  \bibinfo {pages} {4253} (\bibinfo {year} {2014})}\BibitemShut {NoStop}%
\bibitem [{\citenamefont {Havener}\ \emph {et~al.}(2014)\citenamefont
  {Havener}, \citenamefont {Liang}, \citenamefont {Brown}, \citenamefont
  {Yang},\ and\ \citenamefont {Park}}]{Havener2014}%
  \BibitemOpen
  \bibfield  {author} {\bibinfo {author} {\bibfnamefont {R.~W.}\ \bibnamefont
  {Havener}}, \bibinfo {author} {\bibfnamefont {Y.}~\bibnamefont {Liang}},
  \bibinfo {author} {\bibfnamefont {L.}~\bibnamefont {Brown}}, \bibinfo
  {author} {\bibfnamefont {L.}~\bibnamefont {Yang}},\ and\ \bibinfo {author}
  {\bibfnamefont {J.}~\bibnamefont {Park}},\ }\bibfield  {title} {\bibinfo
  {title} {{Van hove singularities and excitonic effects in the optical
  conductivity of twisted bilayer graphene}},\ }\href
  {https://doi.org/10.1021/nl500823k} {\bibfield  {journal} {\bibinfo
  {journal} {Nano Letters}\ }\textbf {\bibinfo {volume} {14}},\ \bibinfo
  {pages} {3353} (\bibinfo {year} {2014})},\ \Eprint
  {https://arxiv.org/abs/1309.2011} {arXiv:1309.2011} \BibitemShut {NoStop}%
\bibitem [{\citenamefont {Wang}\ \emph {et~al.}(2017)\citenamefont {Wang},
  \citenamefont {Kharche}, \citenamefont {{Costa Gir{\~{a}}o}}, \citenamefont
  {Feng}, \citenamefont {M{\"{u}}llen}, \citenamefont {Meunier}, \citenamefont
  {Fasel},\ and\ \citenamefont {Ruffieux}}]{Wang2017}%
  \BibitemOpen
  \bibfield  {author} {\bibinfo {author} {\bibfnamefont {S.}~\bibnamefont
  {Wang}}, \bibinfo {author} {\bibfnamefont {N.}~\bibnamefont {Kharche}},
  \bibinfo {author} {\bibfnamefont {E.}~\bibnamefont {{Costa Gir{\~{a}}o}}},
  \bibinfo {author} {\bibfnamefont {X.}~\bibnamefont {Feng}}, \bibinfo {author}
  {\bibfnamefont {K.}~\bibnamefont {M{\"{u}}llen}}, \bibinfo {author}
  {\bibfnamefont {V.}~\bibnamefont {Meunier}}, \bibinfo {author} {\bibfnamefont
  {R.}~\bibnamefont {Fasel}},\ and\ \bibinfo {author} {\bibfnamefont
  {P.}~\bibnamefont {Ruffieux}},\ }\bibfield  {title} {\bibinfo {title}
  {{Quantum Dots in Graphene Nanoribbons}},\ }\href
  {https://doi.org/10.1021/acs.nanolett.7b01244} {\bibfield  {journal}
  {\bibinfo  {journal} {Nano Letters}\ }\textbf {\bibinfo {volume} {17}},\
  \bibinfo {pages} {4277} (\bibinfo {year} {2017})}\BibitemShut {NoStop}%
\end{thebibliography}%
\end{document}